\documentclass[onecolumn]{article}

\usepackage{cite}
\usepackage{amsmath}
\usepackage{amssymb}
\usepackage{abstract}
 \usepackage{fullpage}
 \usepackage{rotating}
 \usepackage{multicol}
 \usepackage[usenames,dvipsnames]{color}
 \usepackage{tikz}
 \newcommand*{\Line}[3][]{\tikz \draw[#1] #2 -- #3;}
 \usepackage{array}
 \usepackage{caption}
 \usepackage[abs]{overpic}
 \usepackage{rotating}
 \usepackage{multirow}
 \usepackage{mathptm,graphicx,rotate,color}
 \definecolor{myRed}{rgb}{0.901960784313726, 0.027450980392157,  0.164705882352941}
 \definecolor{myGreen}{rgb}{0.427450980392157,   0.752941176470588,   0.282352941176471}
 \definecolor{myBlue}{rgb}{0.258823529411765,   0.309803921568627,   0.643137254901961}
 \definecolor{myCyan}{rgb}{0.490196078431373,   0.803921568627451,   0.862745098039216}
 \definecolor{myMagenta}{rgb}{0.705882352941177,   0.290196078431373,   0.619607843137255}
 \usepackage[colorlinks=true, linkcolor=myBlue, citecolor=myBlue, urlcolor=myBlue]{hyperref}
 \newcommand*{\HEL}{\fontfamily{phv}\selectfont}
 
\title{Constructing a taxonomy of fine-grained human movement and activity motifs 
through social media}
\author{Morgan R. Frank$^{1,2}$, Jake Ryland Williams$^1$, Lewis Mitchell$^{1,3}$,\\ James P. Bagrow$^1$, Peter Sheridan Dodds$^1$, Christopher M. Danforth$^1$\\
{\footnotesize  $^1$Computational Story Lab, Department of Mathematics and Statistics,}\\
{\footnotesize Vermont Complex Systems Center, Vermont Advanced Computing Core,}\\
{\footnotesize University of Vermont, Burlington, Vermont, United States of America }\\ 
\\
{\footnotesize $^2$Current address:}\\
{\footnotesize Massachussetts Institute of Technology,}\\
{\footnotesize Cambridge, MA, USA}\\
\\
{\footnotesize $^3$Current address:}\\
{\footnotesize School of Mathematical Sciences,}\\
{\footnotesize University of Adelaide, Adelaide, Australia}\\
\\
{\footnotesize mrfrank@mit.edu, jake.williams@uvm.edu, lewis.mitchell@uvm.edu,}\\ 
{\footnotesize james.bagrow@uvm.edu, peter.dodds@uvm.edu, chris.danforth@uvm.edu}}

\begin{document}
\maketitle
\begin{multicols}{2}
\begin{abstract}
	\bf Profiting from the emergence of web-scale social data sets, numerous 
recent studies have systematically explored human mobility patterns over 
large populations and large time scales. Relatively little attention, 
however, has been paid to mobility and activity over smaller time-scales, 
such as a day. Here, we use Twitter to identify people's frequently 
visited locations along with their likely activities as a function of time 
of day and day of week, capitalizing on both the content and geolocation 
of messages. We subsequently characterize people's transition pattern 
motifs and demonstrate that spatial information is encoded in word choice.
\end{abstract}
\section{Introduction}
\indent The growth of social media has made possible new insights into human behavior and predictability~\cite{advert,distribute,survey,effect,movies}. Many active areas of research, such as disease spreading~\cite{diseaseSpread,epidemics,forecast,influenza1,influenza2,italy,infect}, traffic forecasting~\cite{traffic,habitual}, urban planning~\cite{urban,agents}, election prediction~\cite{election}, understanding stock market behavior~\cite{flashCrash}, and the spreading of ideas~\cite{predict} have greatly benefited from the recent availability of large-scale social datasets, such as mobile phone data~\cite{movement,bagrow,dynamics,motif,barabasi} and Twitter data~\cite{mystuff,geoHap,hedono,recip}. 
One particular area of interest lies in discerning patterns of human activity, and understanding what these patterns tell us about human life. An example is the ability to identify the typical work day for members of western culture, who comprise most of the English speakers on Twitter~\cite{peer reach}.\\
\indent Previous studies of macroscale human mobility~\cite{mystuff,movement} have been bolstered by investigations into the mechanics of microscale activity. 
Gonz\'alez et al.~\cite{motif} used cellphone data to investigate daily mobility patterns, called ``motifs'', based on cellphone tower reception areas. Their study used data with very fine temporal resolution and coarse spatial resolution.
In this study, we use geolocated Twitter data containing coarse temporal resolution due to subsampling, refined spatial resolution, and, crucially, the lexical content of the Twitter messages~\cite{mystuff}. These differences provide the opportunity for us to use new methods and sensing techniques for understanding human behavior through social data. \\
\indent Twitter is a social media platform where individuals author short messages, called ``tweets", which they can choose to label with their exact location. The combination of message content and spatial location is especially novel in comparison to mobile phone data.
We have seen that changes in the culture of a society may be reflected in large text corpora, such as the Google n-gram dataset~\cite{books,ngram}. Similarly, Twitter can provide large aggregated texts representing the underlying sentiment or well-being of a population of Twitter users~\cite{largeScale,positivity}.
In previous work, we showed that Twitter word usage allows us to demonstrate the characteristics and similarities of American cities~\cite{geoHap}, and to explore the relationship between travel and happiness~\cite{mystuff}.
\begin{figure*}[!t]
	\begin{center}
		$\begin{array}{rl}
			\fbox{\begin{overpic}[scale=.4,trim=0cm 0cm 0cm 0cm,clip]{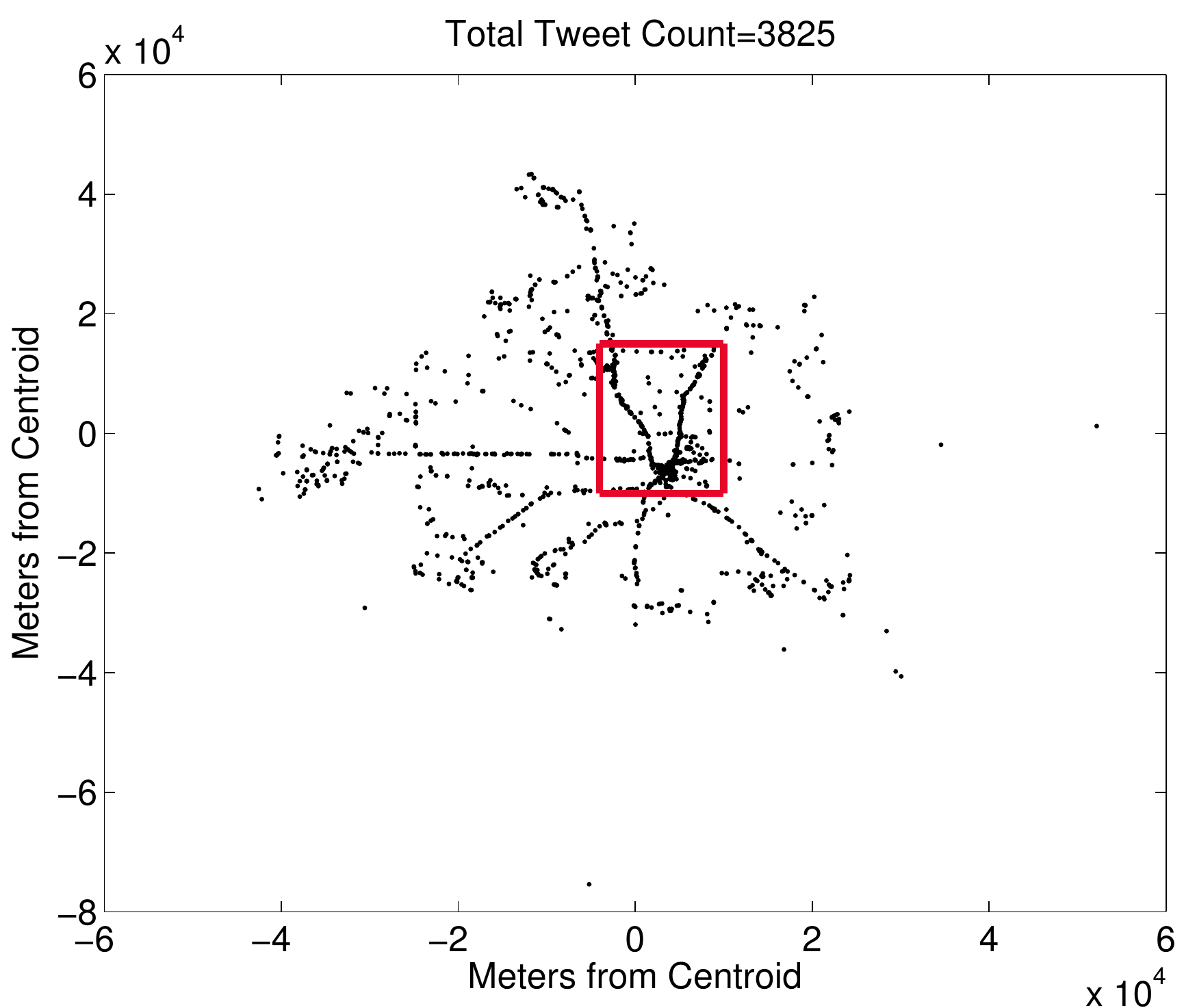}
			\end{overpic}}&
			\hspace{-.35cm}\fbox{\begin{overpic}[scale=.4,trim=0cm -.25cm 0cm 0cm,clip]{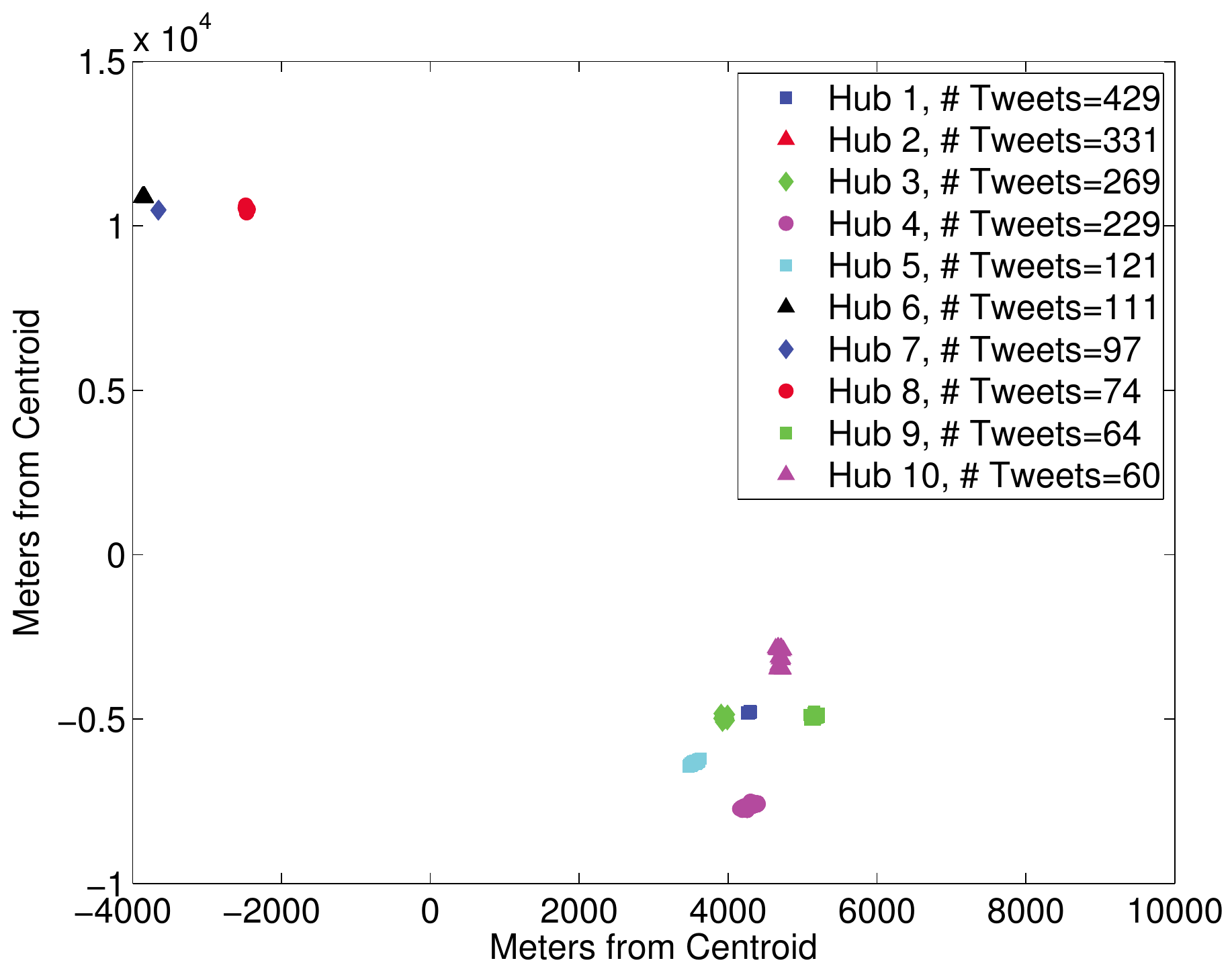}
				\put(-37,7){\Line[myRed, thick]{(0,0)}{(-4.25,2.35)}}
				\put(-37,39){\Line[myRed, thick]{(0,0)}{(-4.2,-1.5)}}
			\end{overpic}}\\
		\end{array}$
	\end{center}
	\vspace{-.5cm}\caption{\bf Tweet hubs for an example prolific Twitter user. (Left) The raw locations for all of the geolocated tweets for this individual. The axes have been scaled to distance in meters from expected location to obfuscate the identity of the individual. 
(Right) A zoomed-in view of a subset of this individual's hubs of activity. Tweet locations are colored by hub, the legend list hubs ranked by frequency. Most hubs are positioned to the southeast of the centroid, while hubs 6, 7, and 8 appear in cluster to the northwest.
}
	\label{exampleHubs}
\end{figure*}
\section{Methods}
\begin{figure*}[t]
	\begin{center}
		\hspace{.75cm}$\begin{array}{cc}
			\hspace{-1.5cm}\begin{overpic}[scale=.35,trim=0cm 0cm 0cm 0cm,clip]{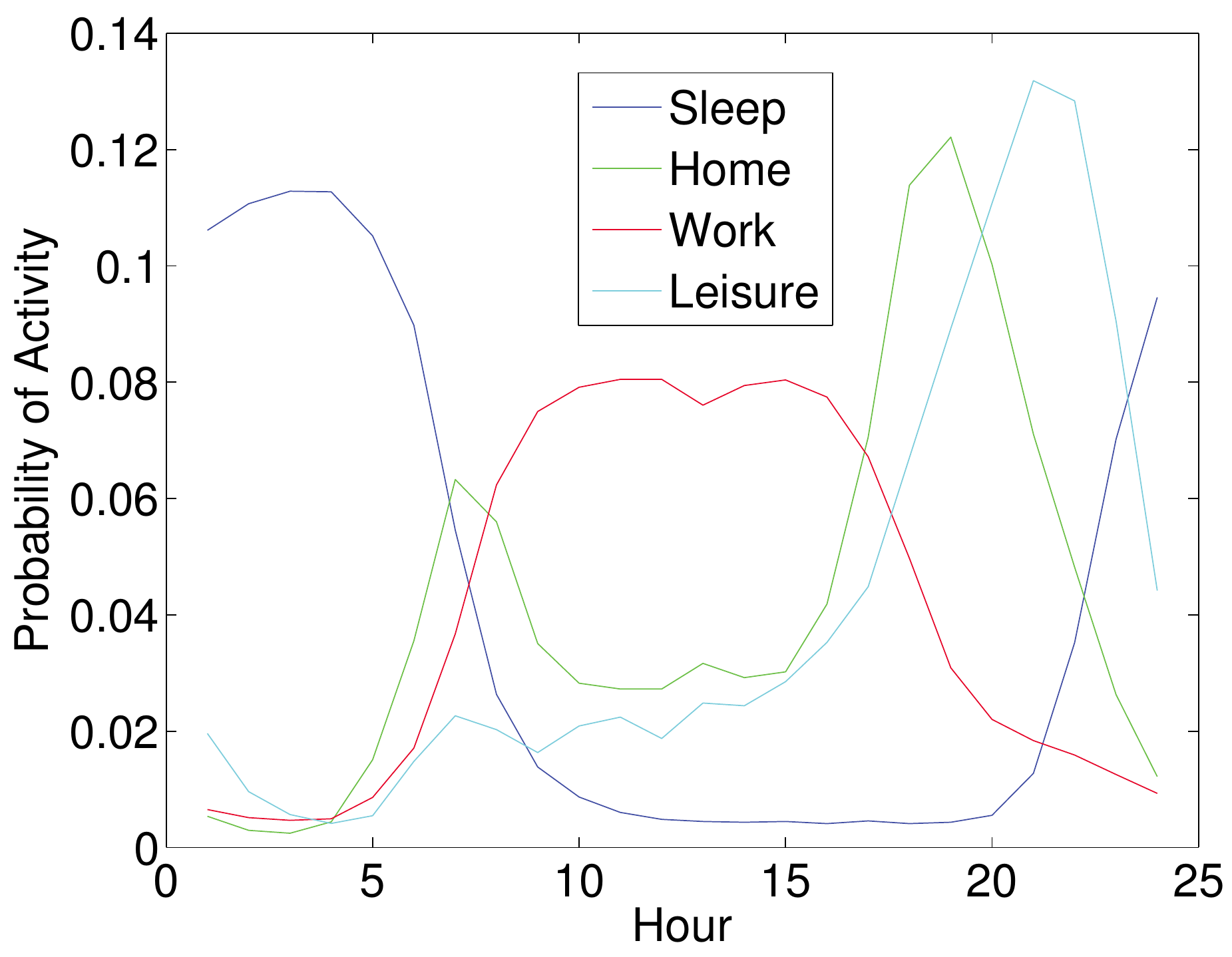}
				\put(15,44){\fbox{A}}
				\put(75,25){$\longrightarrow$}
			\end{overpic}&
			\hspace{2cm}\begin{overpic}[scale=.35,trim=0cm 0cm 0cm 6cm]{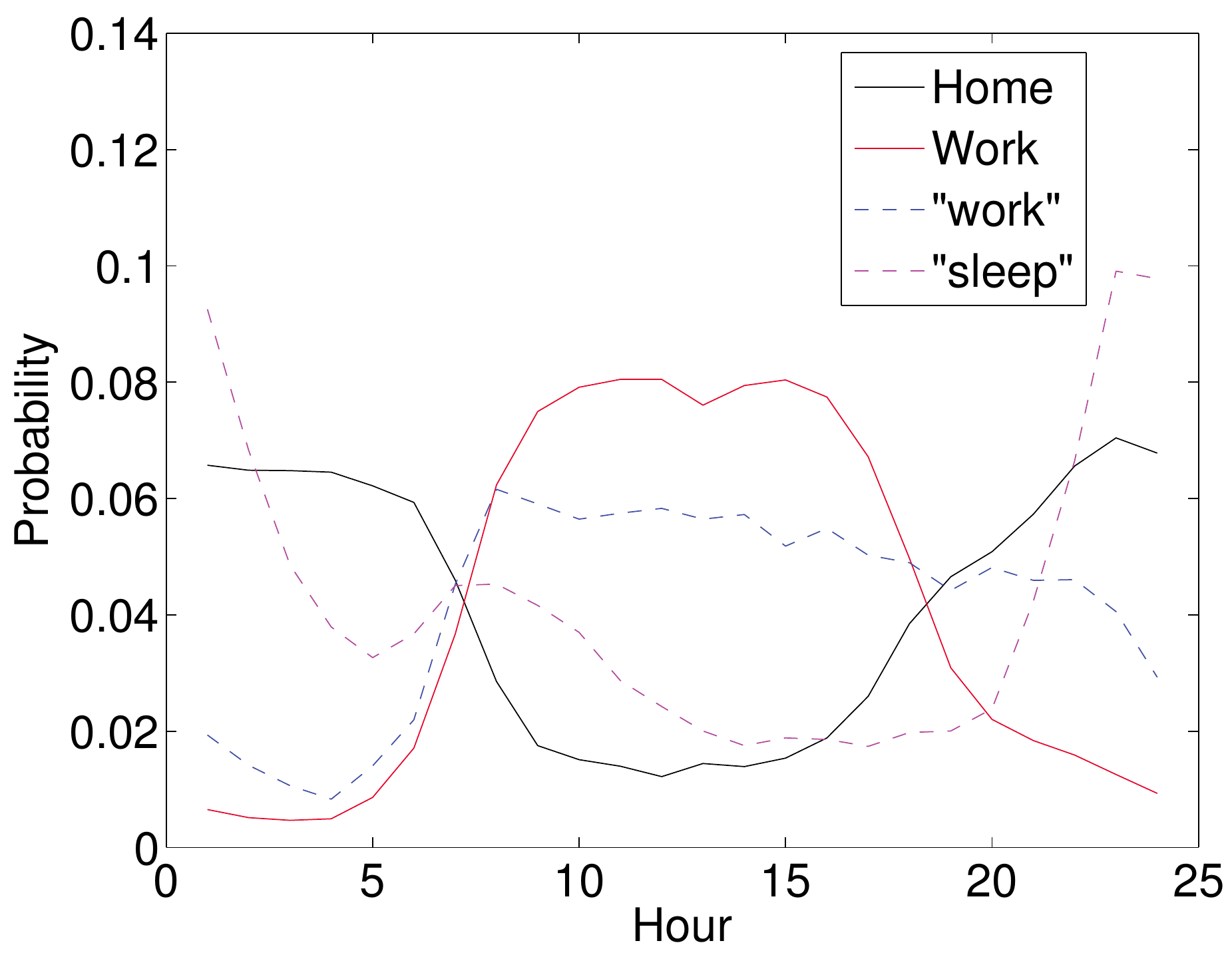}
				\put(15,44){\fbox{B}}
			\end{overpic}\\
		\end{array}$
	\end{center}
	\caption{\bf (A) Probability density functions for each of the four activity classes in the American Time Use Survey~\cite{BLS}. {\bf (B)} We combine the Sleep, Home, and Leisure curves from (A) to obtain a new ``Home" hourly PDF representing non-working activity with a solid black line. The solid \textcolor{myRed}{red} line represents the Work curve. 
The dashed lines represent the hourly PDFs for occurrences of the words ``work" and ``sleep" on twitter during the years 2011 and 2012.
	}
	\label{BLScurves}
\end{figure*}

\begin{figure}[!Ht]
	\begin{center}
		$\begin{array}{cc}
			\fbox{\begin{overpic}[scale=.35,trim=0cm 0cm 0cm 0cm,clip]{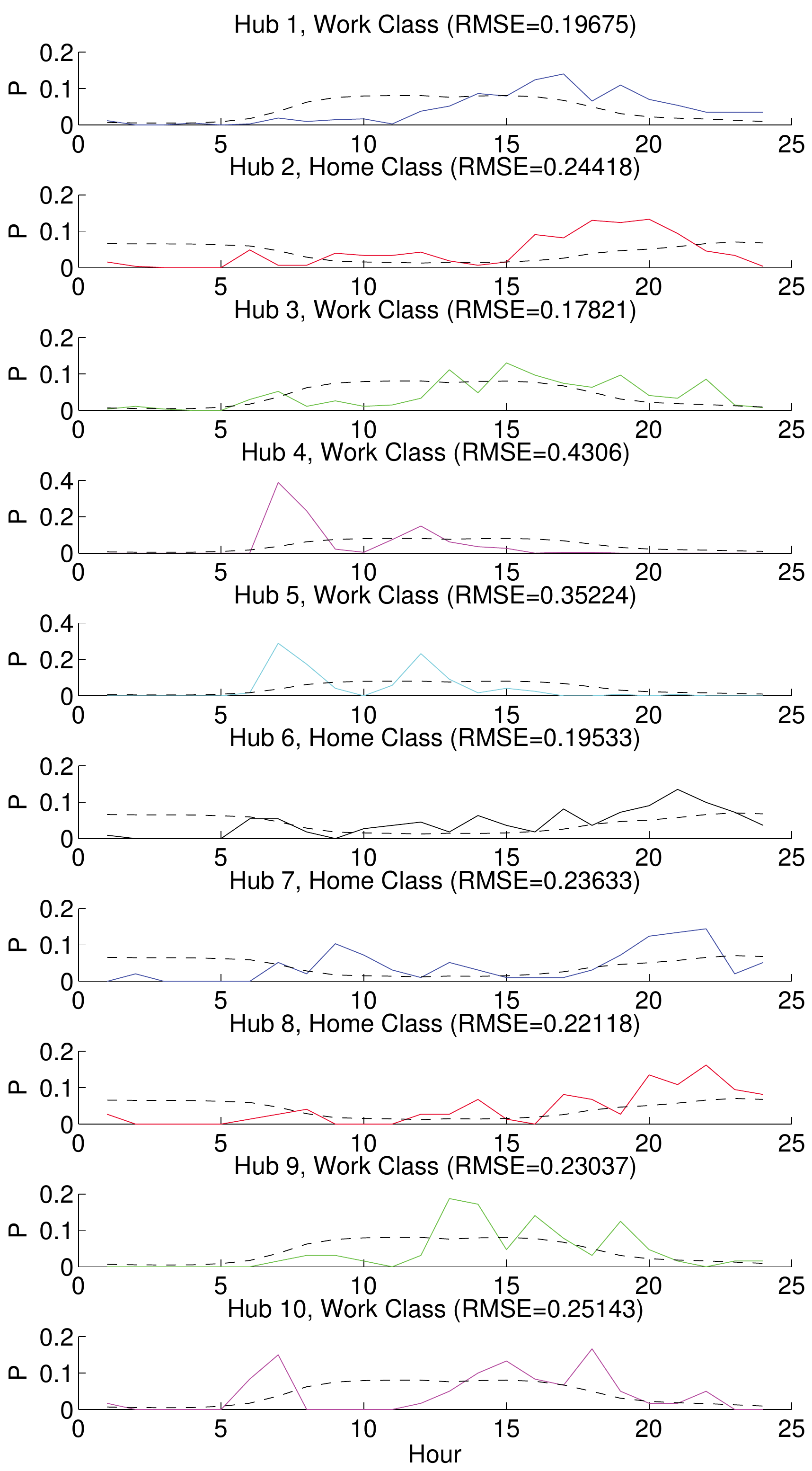}
				\put(7,107){\fbox{\text{A}}}
			\end{overpic}}&
			\hspace{-.35cm}\fbox{\begin{overpic}[scale=.35,trim=0cm 0cm 0cm -.5cm,clip]{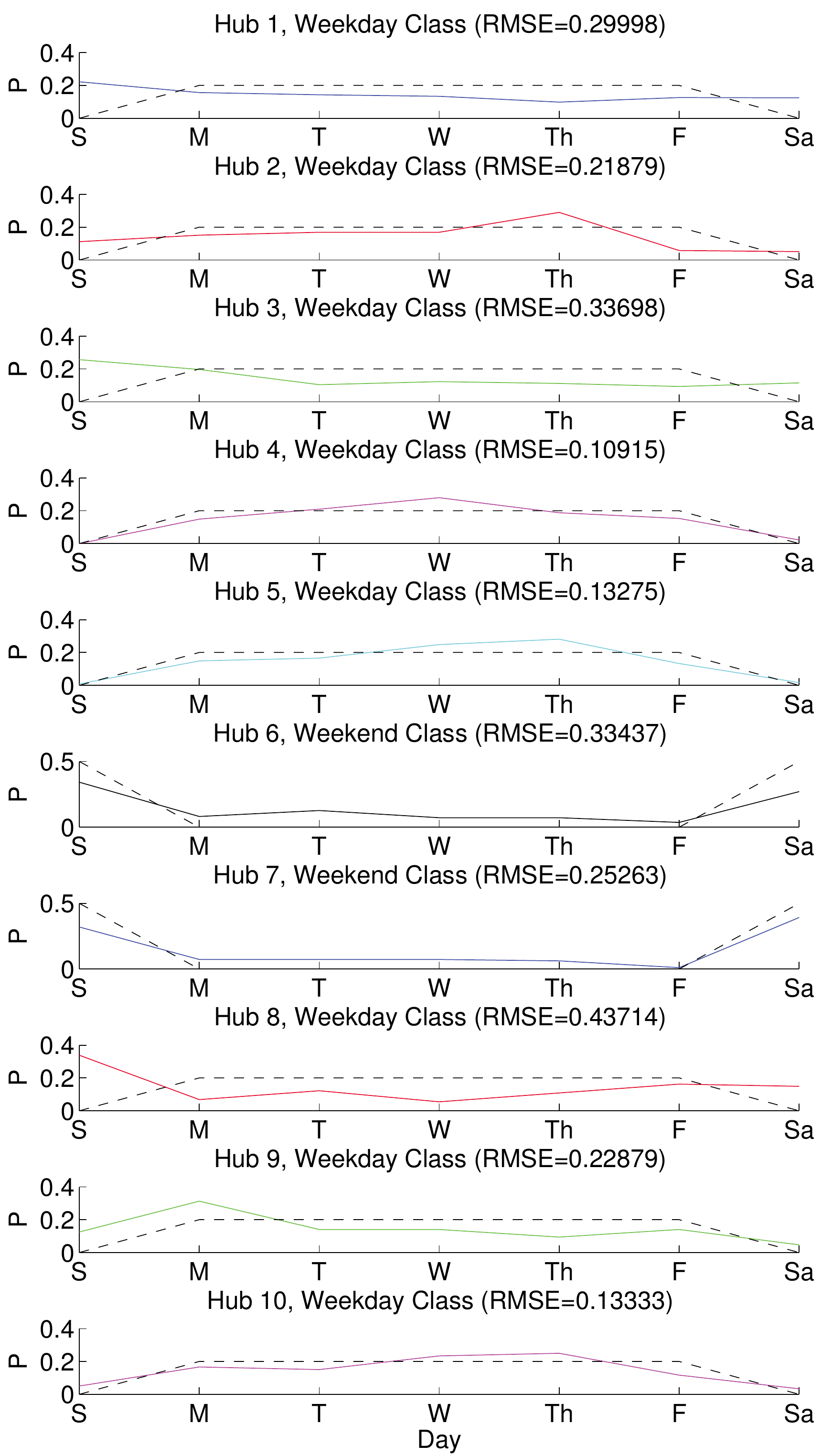}
				\put(7,107){\fbox{\text{B}}}
			\end{overpic}}\\
			\fbox{\begin{overpic}[scale=.4,trim=0cm 0cm .5cm 0cm,clip]{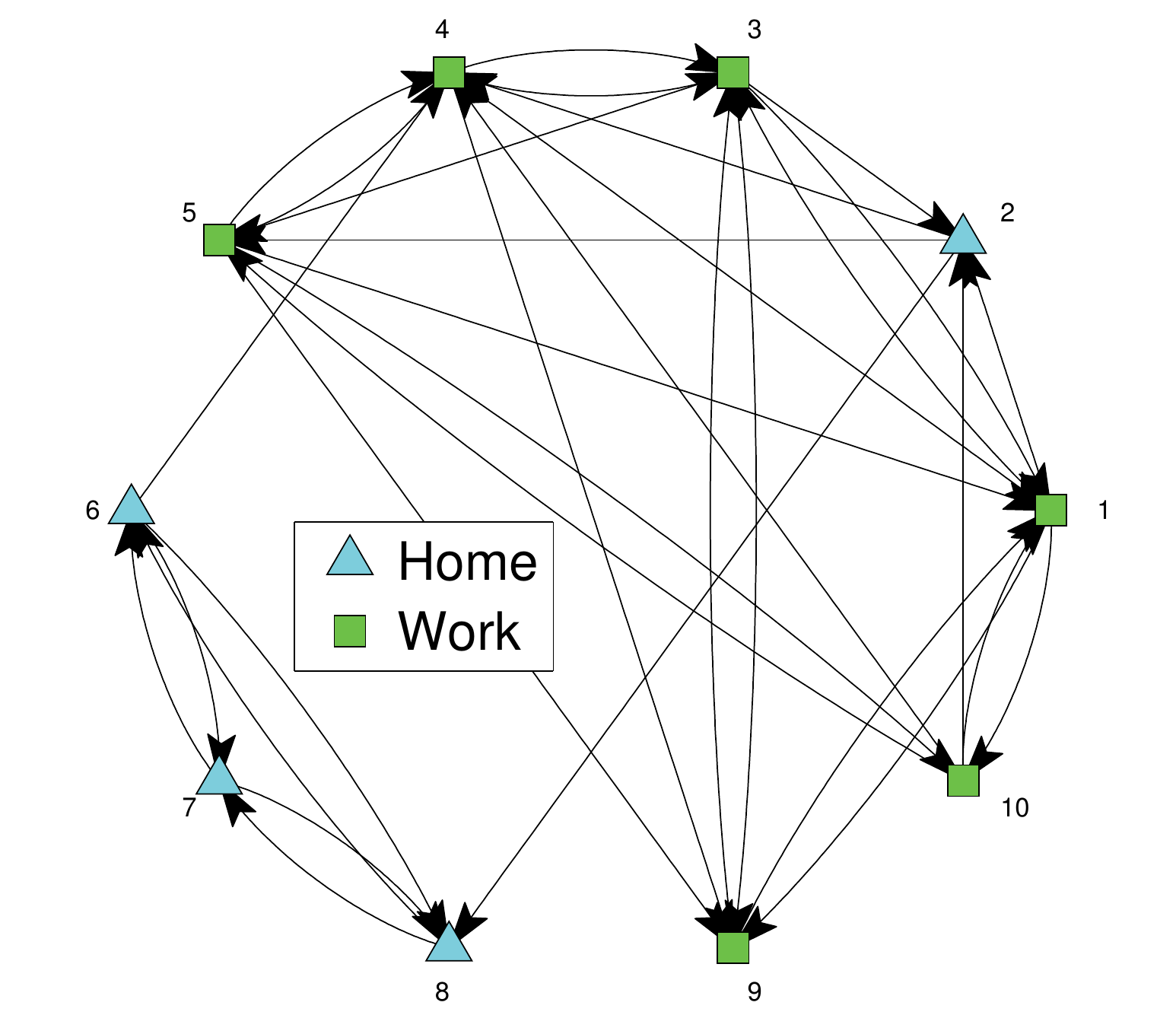}
					\put(2,48){\fbox{\text{C}}}
				\end{overpic}}&
			\hspace{-.35cm}\fbox{\begin{overpic}[scale=.3,trim=-1cm -1cm -1cm -2.5cm]{fig1_hubs-eps-converted-to.pdf}
					\put(1,48){\fbox{\text{D}}}
				\end{overpic}}
			\end{array}$	
	\end{center}
	\vspace{-.5cm}\caption{\bf An example Tweet hub classification and cumulative Twitter motif for the Twitter user from Fig.~\ref{exampleHubs}. {\bf (A)} We classify the ten tweet hubs for this individual according to which of the activity class PDFs the hourly probability density function for that tweet hub is most similar to. For each of these tweet hubs, we provide the hourly PDF for that tweet hub in the color corresponding to the color of that tweet hub in Figure \ref{exampleHubs} in the right plot (note this same plot is provided in ({\bf D}) for reference). The resulting hub classification is provided above each of the PDFs, and the PDF of the activity class is provided as a dashed line for comparison. We perform a similar treatment to obtain the daily classifications shown in panel {\bf(B)}. {\bf (C)} The cumulative Twitter motif for the example individual with the nodes labeled by tweet hub and colored by the activity classification assigned to that tweet hub. Home-activity nodes 6, 7, \& 8 are nearly their own network component, and this separation from the rest of the tweet hubs is reflected in the spatial distance shown in (D).}
	\label{exampleMotif}
\end{figure}

\indent We collected a random ten percent of all Twitter activity through Twitter's garden hose API during 2011 and 2012. Many individuals choose to make their account public, and geolocate their messages from a GPS-enabled device, such as a smartphone. Approximately one percent of all Twitter activity is geolocated, and we collected roughly 150 million geolocated tweets during the time period of this study. Grouping the tweets by their sender, we identify ``prolific" Twitter users as individuals with at least 600 geolocated tweets in our dataset. Given our sampling rate, we expect that these individuals each geolocated roughly 6000 messages during the 2 year period. We aim to investigate the daily pattern of life of these people.
Prolific Twitter users in this study can be found all over the world, however the context analysis we present will reflect English speakers only. The entire user dataset is used to produce context-independent analysis. Prolific Twitter users are filtered to eliminate automated Twitter services, such as weather stations, emergency information, coupon services, job searching services, and even the Big Ben clock in London, England, by investigating repetitive message structure and repetition of key words. We obtained about $1,900$ prolific Twitter users for this study, of which $1,000$ are English speakers.\\
\indent For each individual, we identify frequently visited spatial locations by finding geographic spatial clusters of geolocated tweets. We look for clusters containing at least 50 tweet locations such that the distance between any tweet in the cluster and the next closest tweet location in the cluster is at most 25 meters. We call such clusters ``tweet hubs". Figure \ref{exampleHubs} demonstrates an example individual's raw geolocated tweets along with their tweet hubs. We find that slight perturbations in these parameters do not change our results appreciably.\\
\indent We use results from the American Time Use Survey (ATUS)~\cite{BLS} conducted by the Bureau of Labor Statistics identifying activities by hour of the day in an effort to understand an individual's activity at a tweet hub. Participants in this survey were individuals age 15 and over who were employed full-time on days they worked. These data were collected by averaging survey results conducted in the years 2008 through 2012. The four activity classifications are Sleep, Home, Work, and Leisure, and Fig.~\ref{BLScurves}A exhibits the hourly probability density functions (hourly PDFs) for each of those classes. The Sleep and Work classes are each clearly distinct from the hourly PDFs of any other activity class. Also, the hourly PDFs for the Home and Leisure classes are similar in the evening hours, but the curve representing the Home class exhibits a small peak in probability in the morning hours that is absent for the Leisure hourly PDF. Our preliminary investigations demonstrated that there were insufficient numbers of tweet hubs of the Sleep activity classification to merit meaningful results, and a contextual analysis of the vocabularies for tweet messages contained in Home or Leisure tweet hubs showed surprising similarities making them indistinguishable (details on these analyses below). Therefore, we combine the Sleep, Home, and Leisure activity curves shown in Figure \ref{BLScurves}A to produce the new Home hourly PDF shown in Figure \ref{BLScurves}B, which represents the union of the three separate activity classifications.\\
\indent We classify each tweet hub as representing either a Home or Work activity by measuring the root mean square error (RMSE) between the tweet hub and ATUS PDFs.
This method provides circumstantial evidence to explain why an individual frequents a particular area represented by a tweet hub. Figures \ref{exampleMotif}A \& \ref{exampleMotif}B provide an example of this activity classification for tweet hubs from the individual in Fig.~\ref{exampleHubs}. \\
\indent Gonz\'alez et al.~\cite{motif} viewed daily mobility patterns for individuals as networks based on their cellphone tower reception: nodes represent tower reception areas and directed edges represent travel between two reception areas. Inspired by this idea, we construct ``cumulative Twitter motifs" as networks where each tweet hub of a given individual is represented as a node and weighted directed edges exists between node $A$ and node $B$ if the individual had a geolocated tweet contained in tweet hub $A$ and then authored a tweet from hub $B$ within a two hour time period. We call this a ``transition" from tweet hub A to tweet hub B. We ignore transitions that do not occur at least five times in our dataset. This construction does not achieve as precise a measurement of daily mobility for the individuals as in~\cite{motif}, but nonetheless provides good approximations for the individuals' mobility patterns. This difference of methods is necessary because of the discrete temporal nature of the Twitter data. Our method of constructing cumulative Twitter motifs may result in networks with separate components (i.e., transitioning between locations in each of the components takes longer than two hours, or there is no data to provide evidence for a transition); we treat each of these components separately when investigating human mobility patterns. We present an example motif in Fig.~\ref{exampleMotif}C.\\
%
\indent The tweet message content provides a unique insight into the users' lives that is absent in other human mobility datasets. Previous studies~\cite{mystuff,geoHap,hedono,recip,ngram,books,largeScale,positivity} have demonstrated the usefulness of examining the vocabulary of large collections of tweets and other text sources. We demonstrate the predictive power of words used at a tweet hub by comparing binary decision trees~\cite{tree1,tree2,tree3} and Bayesian document classifiers~\cite{bayes1,bayes2} as they predict the 
\end{multicols}
\clearpage
\clearpage
\begin{multicols}{2}
\noindent  activity classifications of tweet hubs in our dataset using word counts for the 50,000 most frequently used words in the English language according to Google Books, music lyrics, New York Times articles, and Twitter messages~\cite{hedono}. For each of these words, we obtain a word count representing the number of occurrences of that word in a text. As a baseline, we will note the improvement in the fraction of correct activity class predictions yielded from a null predictor, which is obtained by randomly permuting the true tweet hub classifications in our dataset and noting which classifications remain correct. The probability of a correct prediction under this process is given by
\begin{equation}
	P(\text{correct})=P(\text{Home})^2+P(\text{Work})^2,
	\label{analytic}
\end{equation}
where $P(\text{Home})=0.65$ and $P(\text{Work})=0.35$ according to tweet hubs taken from English speaking Twitter users. This results in a binomial distribution with a mean of $P(\text{correct})=0.545$ and a variance of $P(\text{correct})(1-P(\text{correct}))=0.248$. We could have naively assumed each activity classification is equally likely to occur, but the comparison to the more sophisticated null model is more realistic. \\
\indent We randomly select half of the tweet hubs in our dataset as training data for a binary decision tree. The remaining tweet hubs are our validation dataset. To construct the binary decision tree, we first examine every possible binary split on each of the 10,000 word counts. We select the split that minimizes Gini's index~\cite{gini1,gini2}, which measures the diversity of the resulting classification predictions. We represent the word count and the binary splitting criterion as a node on the decision tree with two child nodes. We recursively repeat this procedure for the children nodes until a node is found to yield predictions of only one class type according to the training data. Nodes with no children are called ``leaf nodes". It is often found that binary decision trees over-fit the training data during construction, and improvements in prediction on validation data can be obtained by systematically removing, or ``pruning", leaf nodes and nodes that are below a certain depth. We select the level of pruning that maximizes the fraction of correct predictions on the validation data.\\
\indent In training the Bayesian document classifiers, we randomly select half of the tweet hubs to be training data, and we leave the remaining tweet hubs for validation. We restrict to $N=1,000$ most frequently occurring words in tweet hubs from the training dataset. For each activity classification, we combine and normalize the word counts from tweet hubs of that class in the training data to obtain a likelihood of the word occurring given the activity class. We denote this by $P(w_i \mid c)$ where $w_i$ is the $i$th word for $i\in\{1,\dots,1000\}$, and $c\in\{\text{Home},\text{Work}\}$ is the activity classification. Thus, given a tweet hub, $H$, the posterior probability that $H$ is of class $c_H=c$ is given by 
\begin{equation}
	P(c_H = c)=P(c)\cdot \prod_{i=1}^N P(w_i\mid c)^{f_H(w_i)},
	\label{bayesClass}
\end{equation}
where $P(c)$ is the probability of activity classification $c$ and $f_H(w_i)$ is the frequency of word $w_i$ occurring in tweet hub $H$. The activity classification with greatest posterior probability is selected as the predicted activity classification for $H$.\\
\indent Calculating the posterior probabilities in Eq.~\ref{bayesClass} directly can lead to large computer roundoff errors. Thus, we instead consider the log probability when implementing the classifier. This allows us to frame the calculation in terms of minimizing surprisal as opposed to maximizing likelihood. Furthermore, since the two activity classifications in our dataset exist in roughly equal proportions, the majority of variation in the posterior probabilities is due to the likelihood function and the performance of the Bayesian document classifier is not appreciably diminished by assuming a flat prior probability distribution, i.e., $P(\text{Home})=P(\text{Work})$. Accordingly, Eq.~\ref{bayesClass} becomes
\begin{equation}
	-\log P(c_H=c)=-\log(2)-\sum_{i=1}^N f_H(w_i)\cdot\log P(w_i\mid c).
	\label{bayes2}
\end{equation}
\indent Equation (\ref{bayes2}) provides an interesting opportunity to determine which words discriminate most strongly between the two activity classifications. We examine the difference in the logs of the two posterior probabilities, $P_{\mathrm{comp}}$, for a tweet hub $H$, to find
\begin{equation}
	P_{\mathrm{comp}}=\sum_{i=1}^N f_H(w_i)\Big(\log P(w_i\mid \text{Home})-\log P(w_i\mid \text{Work})\Big).
	\label{bayes3}
\end{equation}
We can rank words by their contribution to the total by examining the sum term by term. This ranking allows us to see the contextual differences in the vocabularies of the Home and Work activity classifications.\\
\section{Results}
\begin{figure*}[!tt]
	\begin{center}
		$\begin{array}{ccc}
			\begin{overpic}[width = .31\textwidth,trim=0cm 0cm 0cm 0cm,clip]{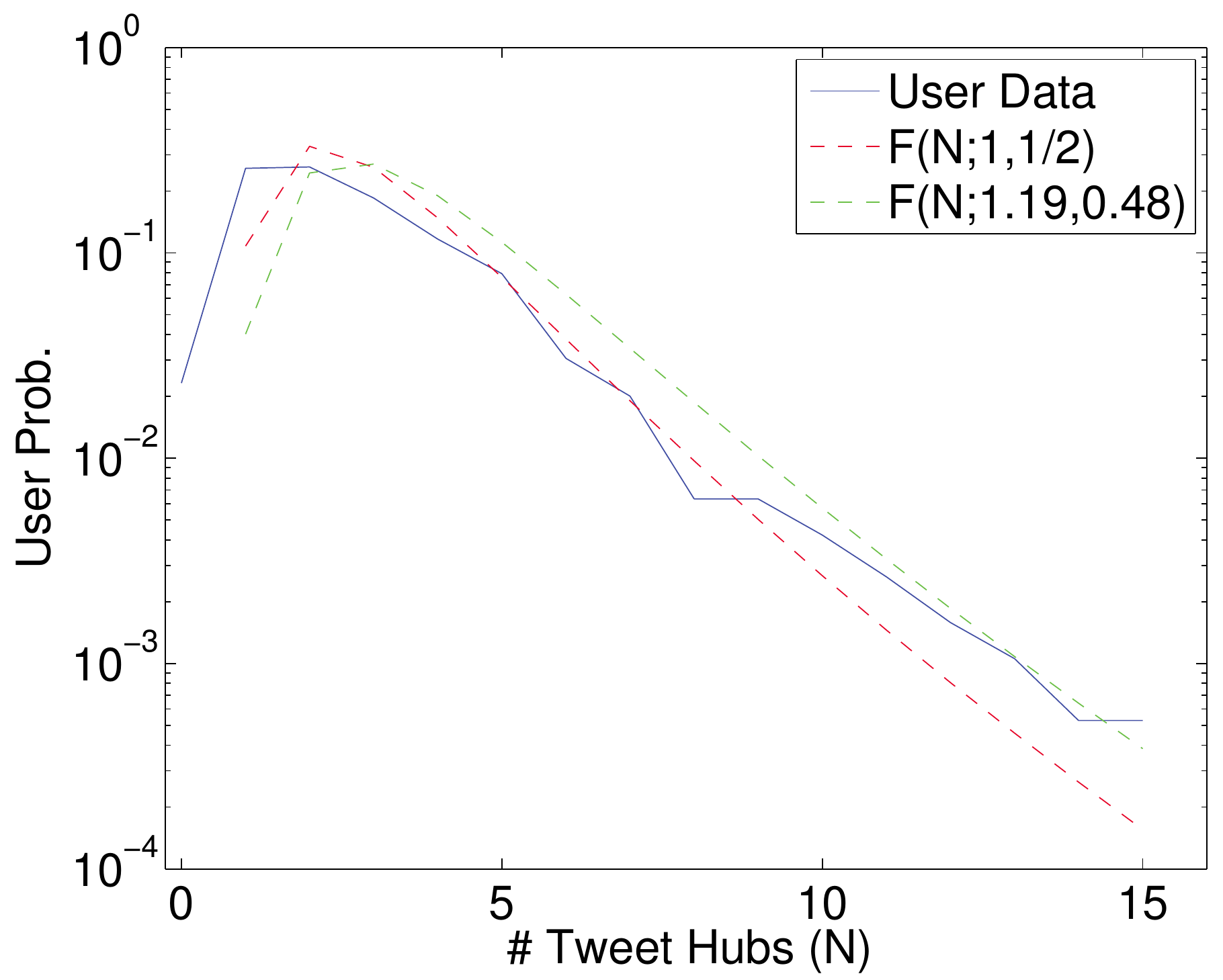}
					\put(15,25){\fbox{\text{A}}}
			\end{overpic}&
			\begin{overpic}[width = .31\textwidth,height=4cm,trim=0cm -.5cm 0cm 0cm]{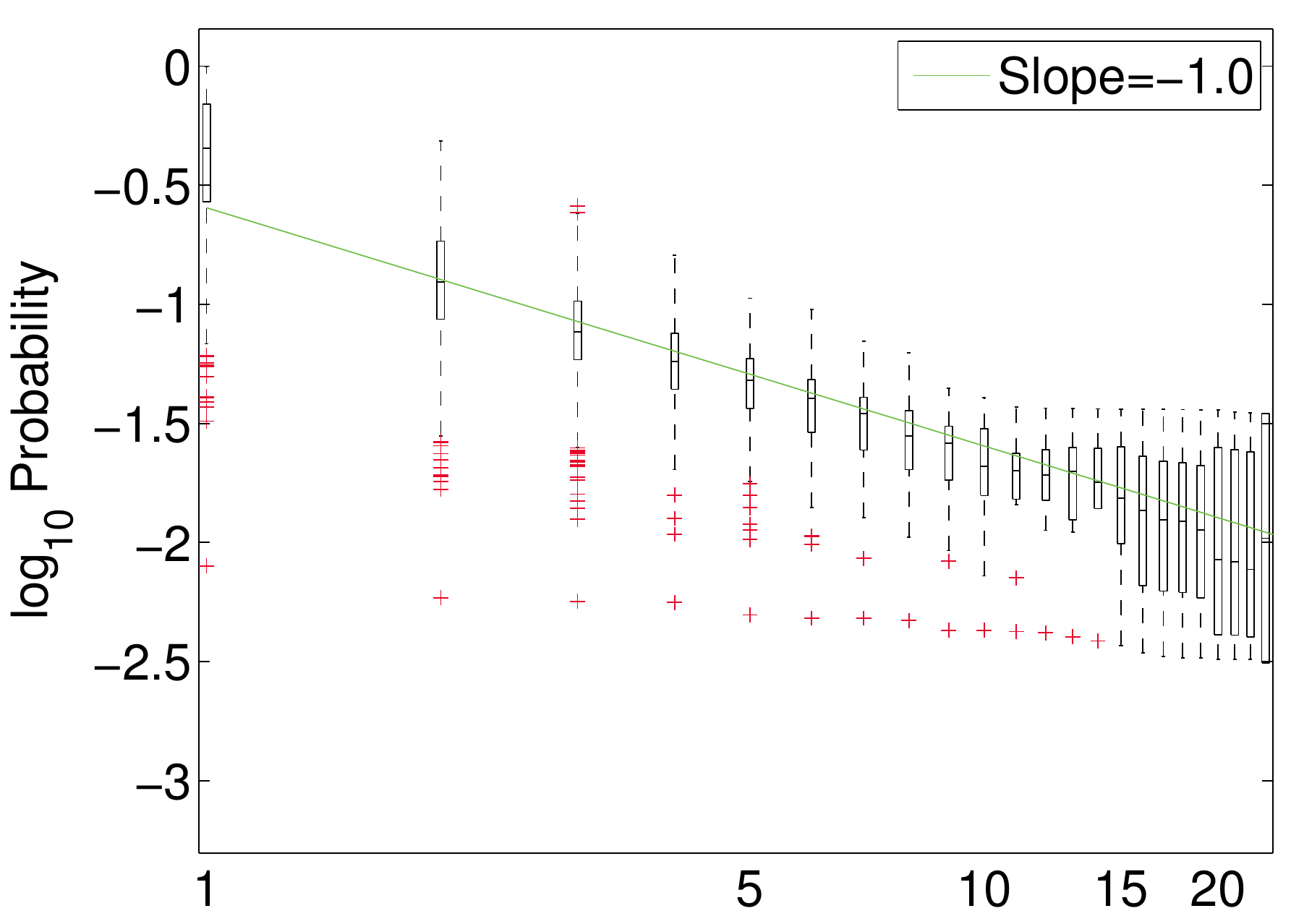}
				\put(10,34){\fbox{\text{B}}}
				\put(25,.5){\text{\fontsize{2mm}{.5em}{\HEL Hub Rank}}}
			\end{overpic}&
			\begin{overpic}[width = .31\textwidth,trim=0cm 0cm 0cm 0cm,clip]{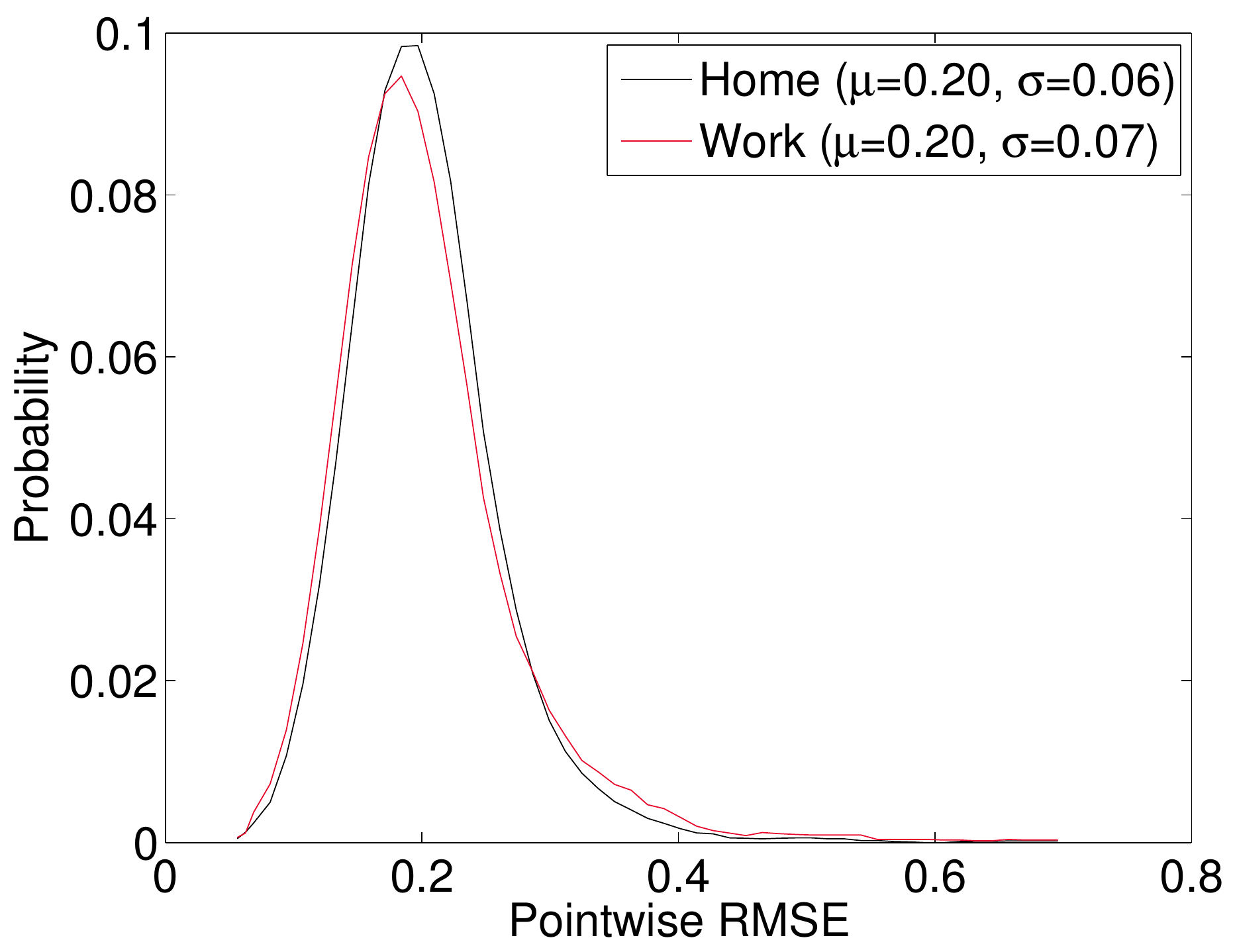}
				\put(9,33.5){\fbox{\text{C}}}
			\end{overpic}
		\end{array}$
	\end{center}
	\vspace{-.5cm}\caption{\bf {\bf (A)} A semi-logarithmic plot demonstrating the lognormal relationship between the number of tweet hubs ($N$) to the probability of individuals with that many tweet hubs. As was found in~\cite{motif}, we see that the relationship is well modeled by $F(N)=\exp\left(-\frac{(\ln N-\mu)^2}{2\sigma^2}\right)/(\sqrt{2\pi}N\sigma)$ with $\mu=1$ and $\sigma=1/2$ (\textcolor{myRed}{red dashed line}); however, we find that the model fits our data slightly better in terms of least squares with $\mu\approx1.19$ and $\sigma\approx0.48$ (\textcolor{myGreen}{green dashed line}). {\bf (B)} After ranking the tweet hubs of each individual by tweet count, we compare tweet hub rank to the probability of the individual of that tweet hub tweeting from that tweet hub. For each hub rank, we provide a box-plot of the probabilities of tweets occurring in tweet hubs of that rank. Outliers are represented by \textcolor{myRed}{red crosses}. We provide a \textcolor{myGreen}{green solid line} with slope $-1$ as a guide for the eye. The box-plots for lower ranking tweet hubs represent many more data points than the box-plots for the higher ranking tweet hubs because only a few individuals were found to have more than 10 tweet hubs (see {\bf A}). {\bf (C)} The distributions for RMSE for the tweet hub hourly PDFs compared to the activity PDF they were most similar to. 
}
	\label{hubCount}
\end{figure*}

\begin{figure*}[!tt]
	\begin{center}
		$\begin{array}{|c|c|c|}
			\hline
			\begin{overpic}[width=.3\textwidth,trim=0cm 7cm 0cm 5cm,clip]{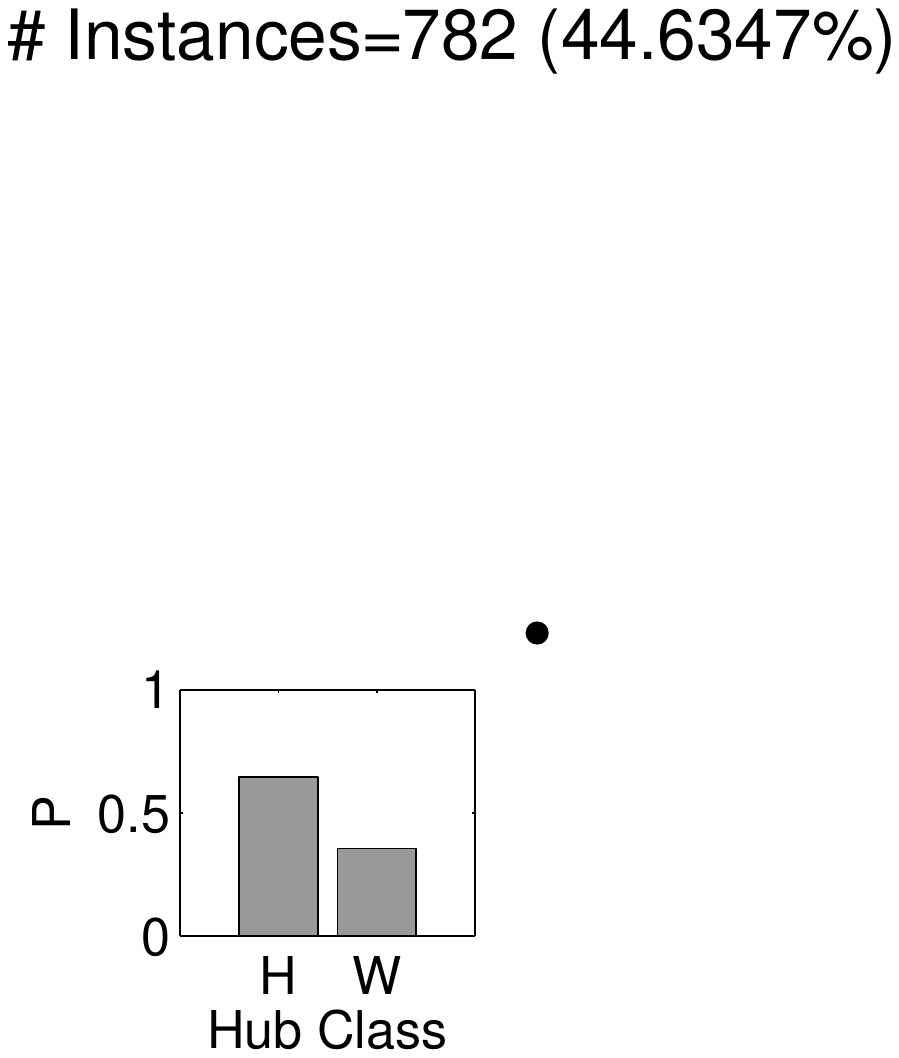}
				\put(37,30){\fbox{\text{Rank 1}}}
			\end{overpic}&
			\begin{overpic}[width=.3\textwidth,trim=0cm 7cm 0cm 5cm,clip]{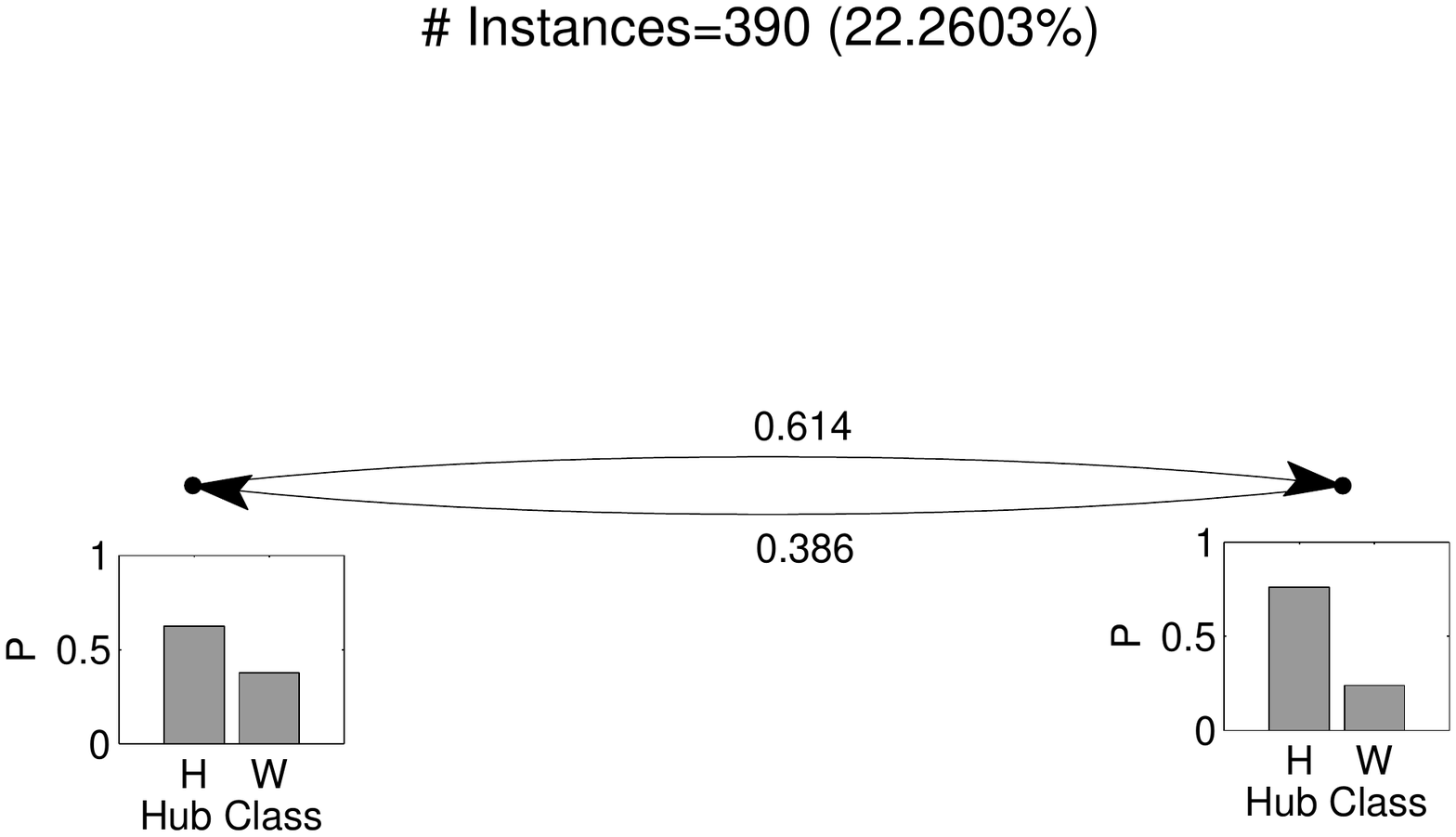}
				\put(37,30){\fbox{\text{Rank 2}}}
			\end{overpic}&			
			\begin{overpic}[width=.3\textwidth,trim=0cm 7cm 0cm 5cm,clip]{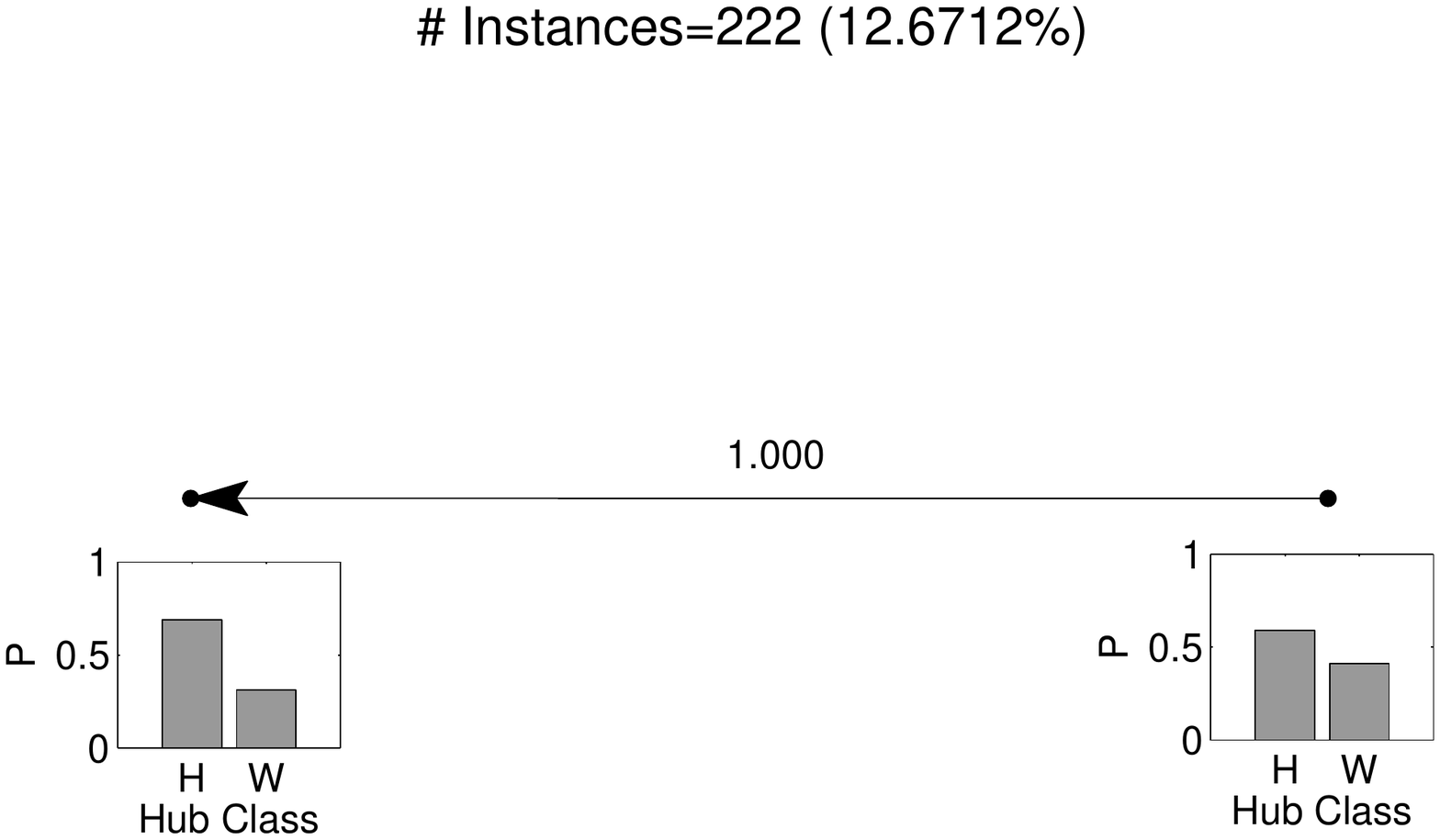}
				\put(37,30){\fbox{\text{Rank 3}}}
			\end{overpic}\\ \hline			
			\begin{overpic}[width=.3\textwidth,trim=0cm 7cm 0cm 5cm,clip]{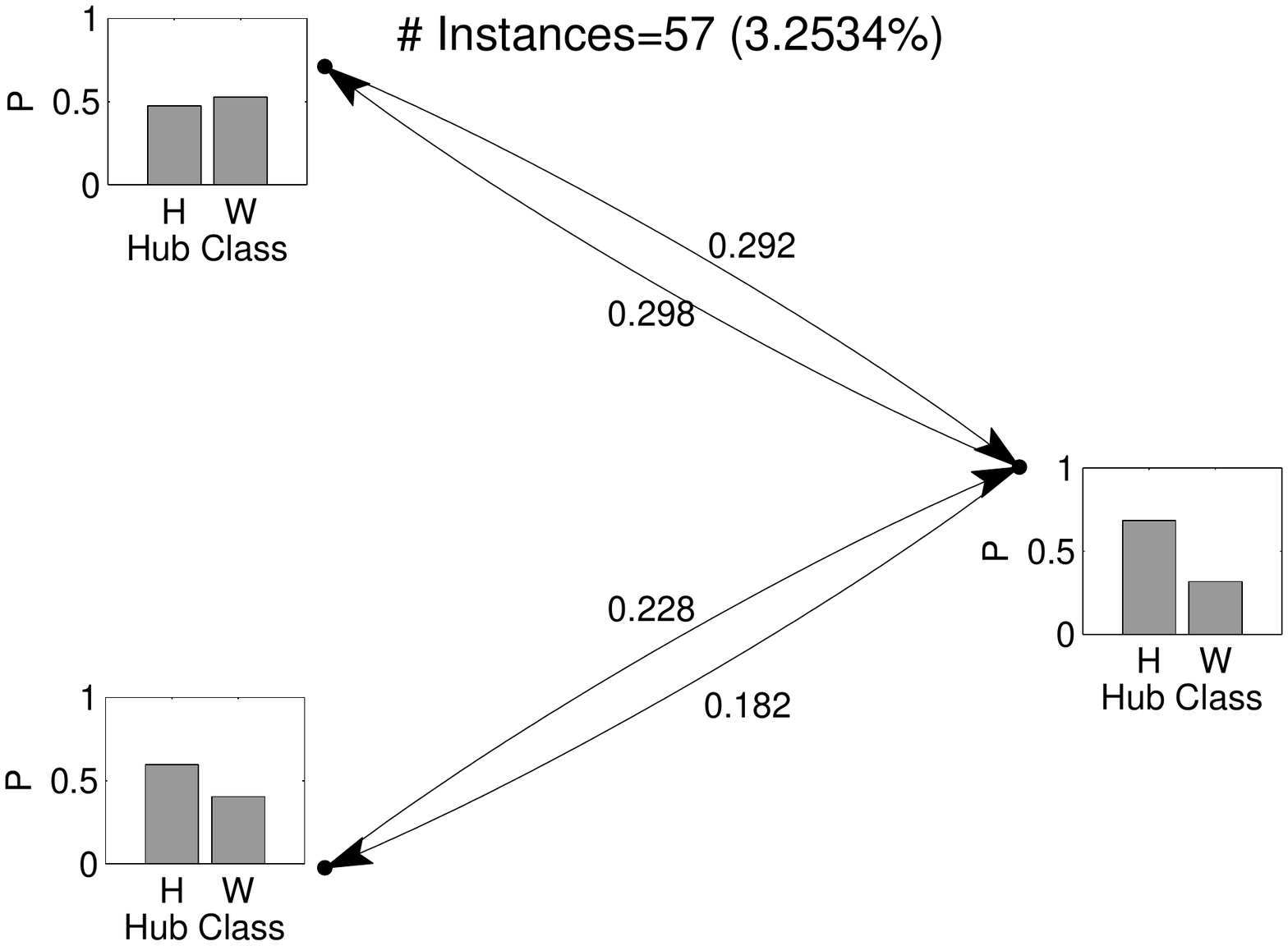}
				\put(37,30){\fbox{\text{Rank 4}}}
			\end{overpic}&
			\begin{overpic}[width=.3\textwidth,trim=0cm 7cm 0cm 5cm,clip]{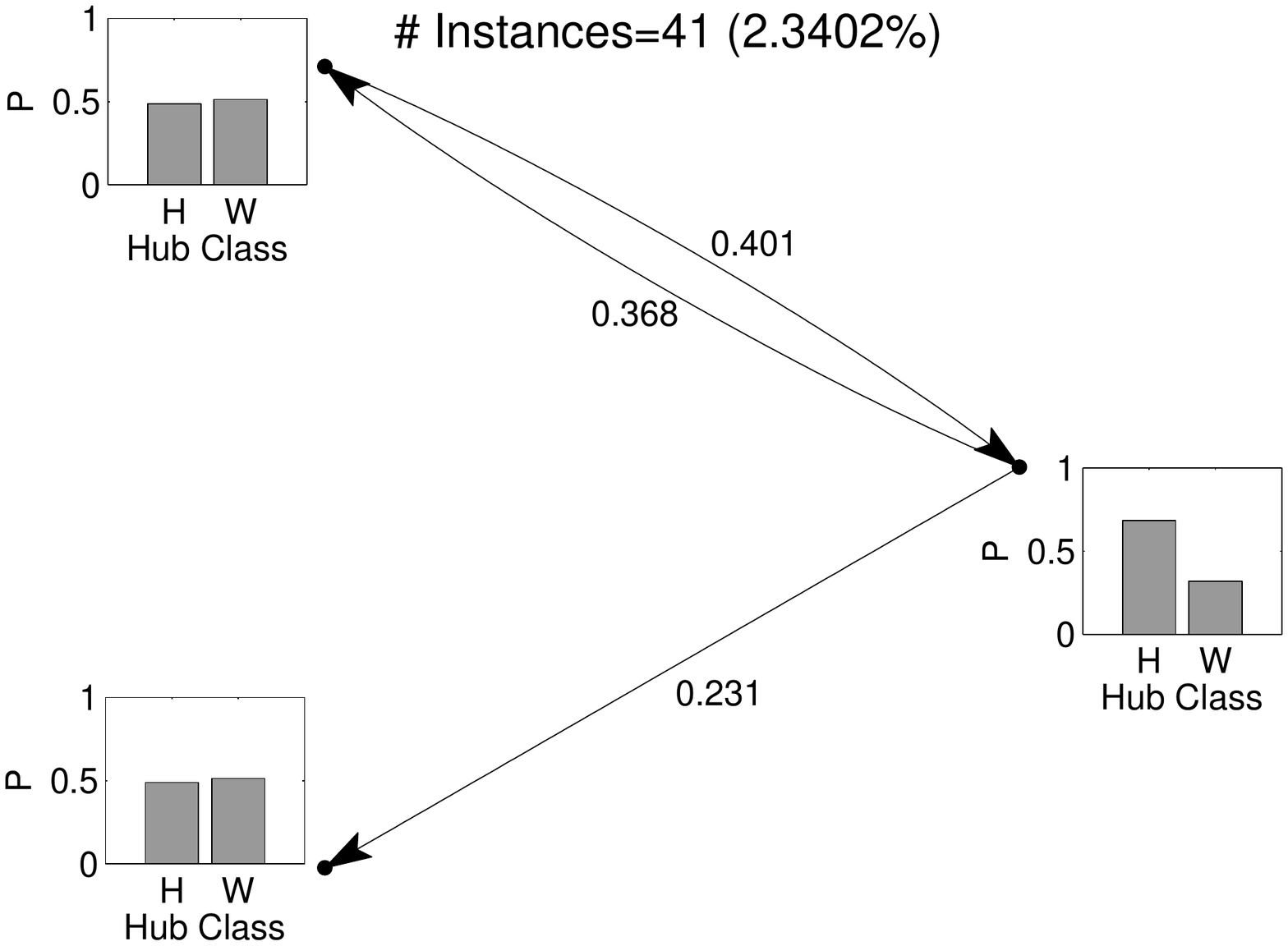}
				\put(37,30){\fbox{\text{Rank 5}}}
			\end{overpic}&
			\begin{overpic}[width=.3\textwidth,trim=0cm 7cm 0cm 5cm,clip]{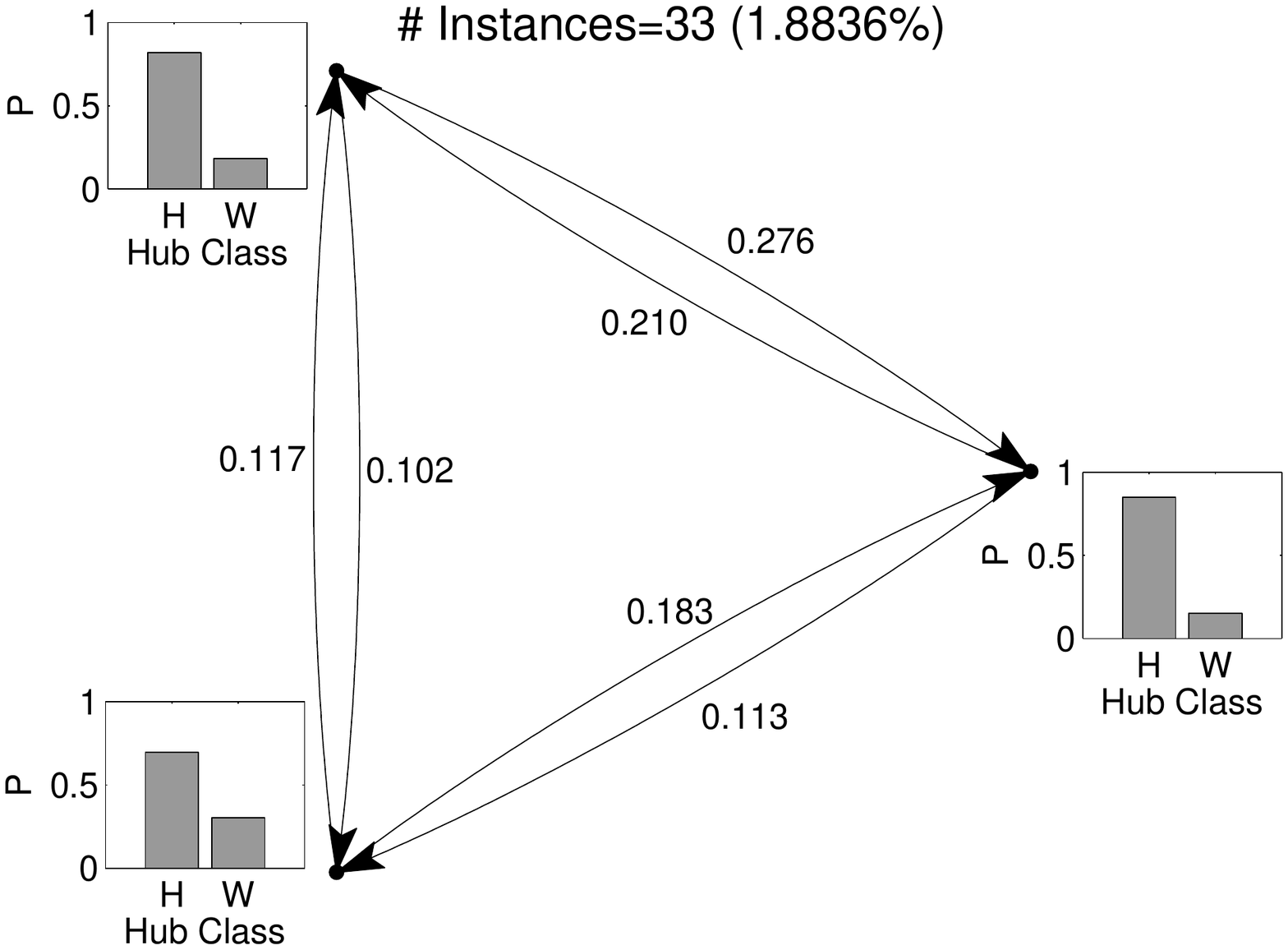}
				\put(37,30){\fbox{\text{Rank 6}}}
			\end{overpic}\\ \hline					
			\begin{overpic}[width=.3\textwidth,trim=0cm 7cm 0cm 5cmm,clip]{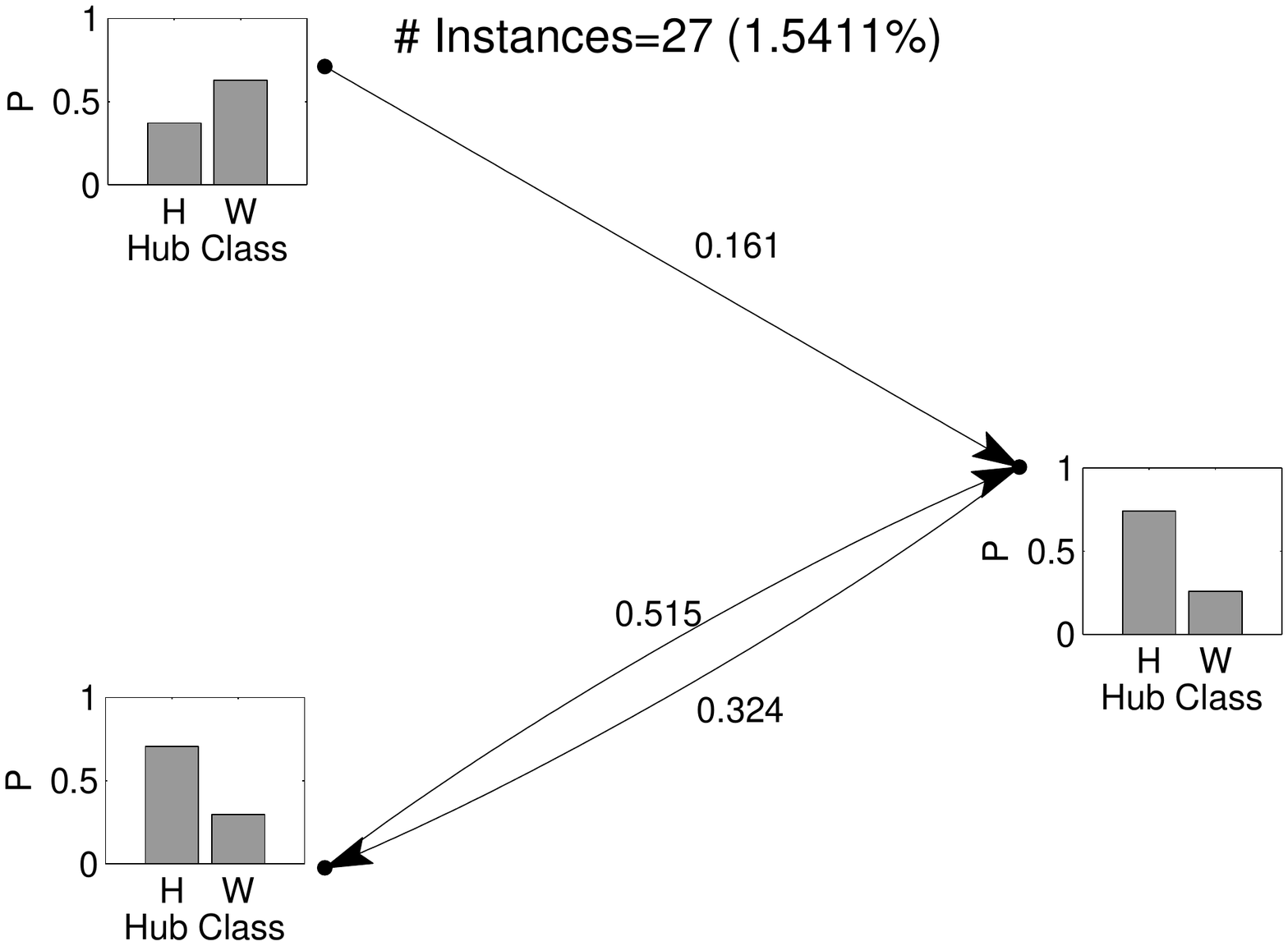}
				\put(37,30){\fbox{\text{Rank 7}}}
			\end{overpic}&
			\begin{overpic}[width=.3\textwidth,trim=0cm 7cm 0cm 5cm,clip]{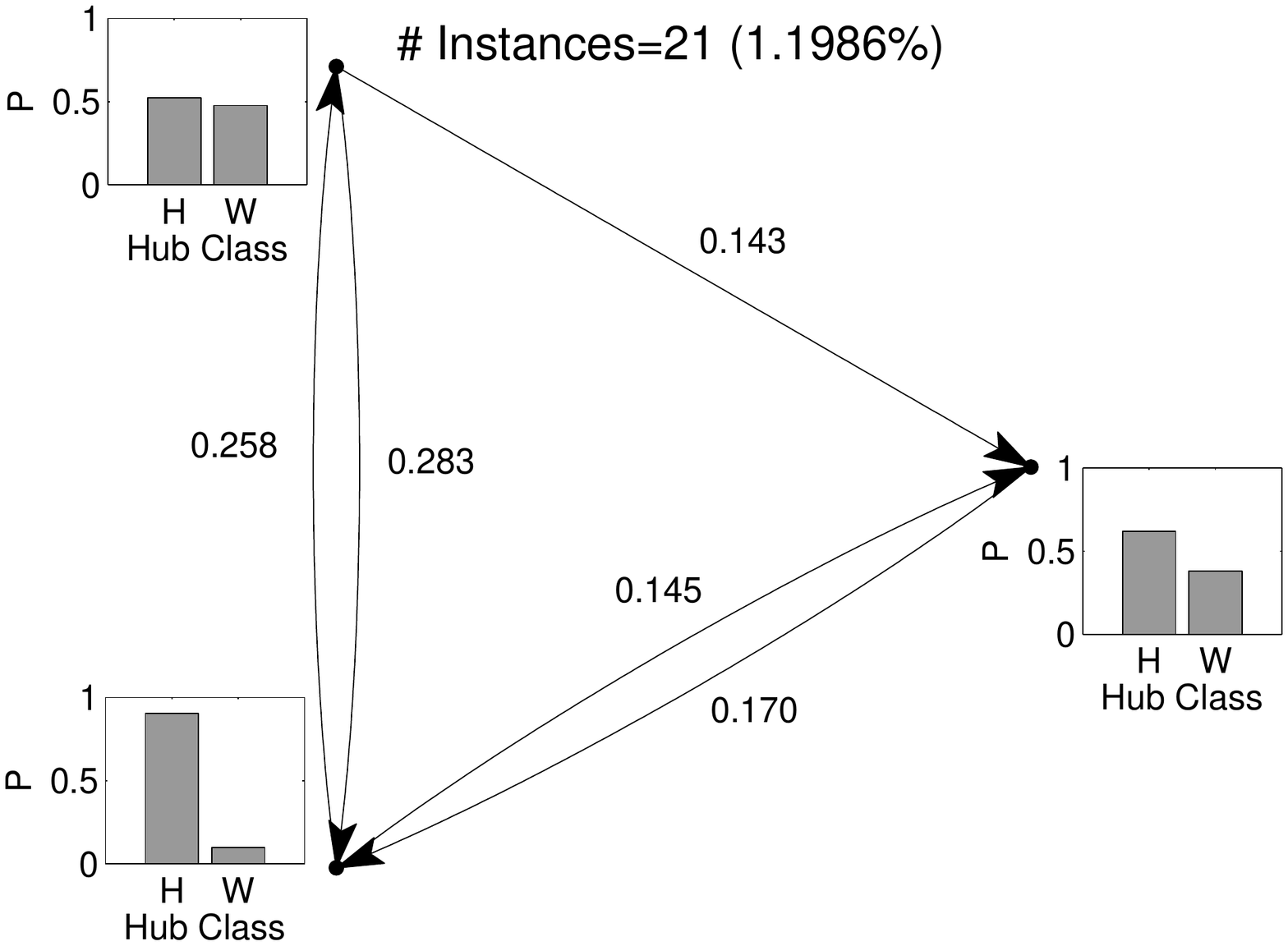}
				\put(37,30){\fbox{\text{Rank 8}}}
			\end{overpic}&
			\begin{overpic}[width=.3\textwidth,trim=1cm 6cm 2cm 7cm,clip]{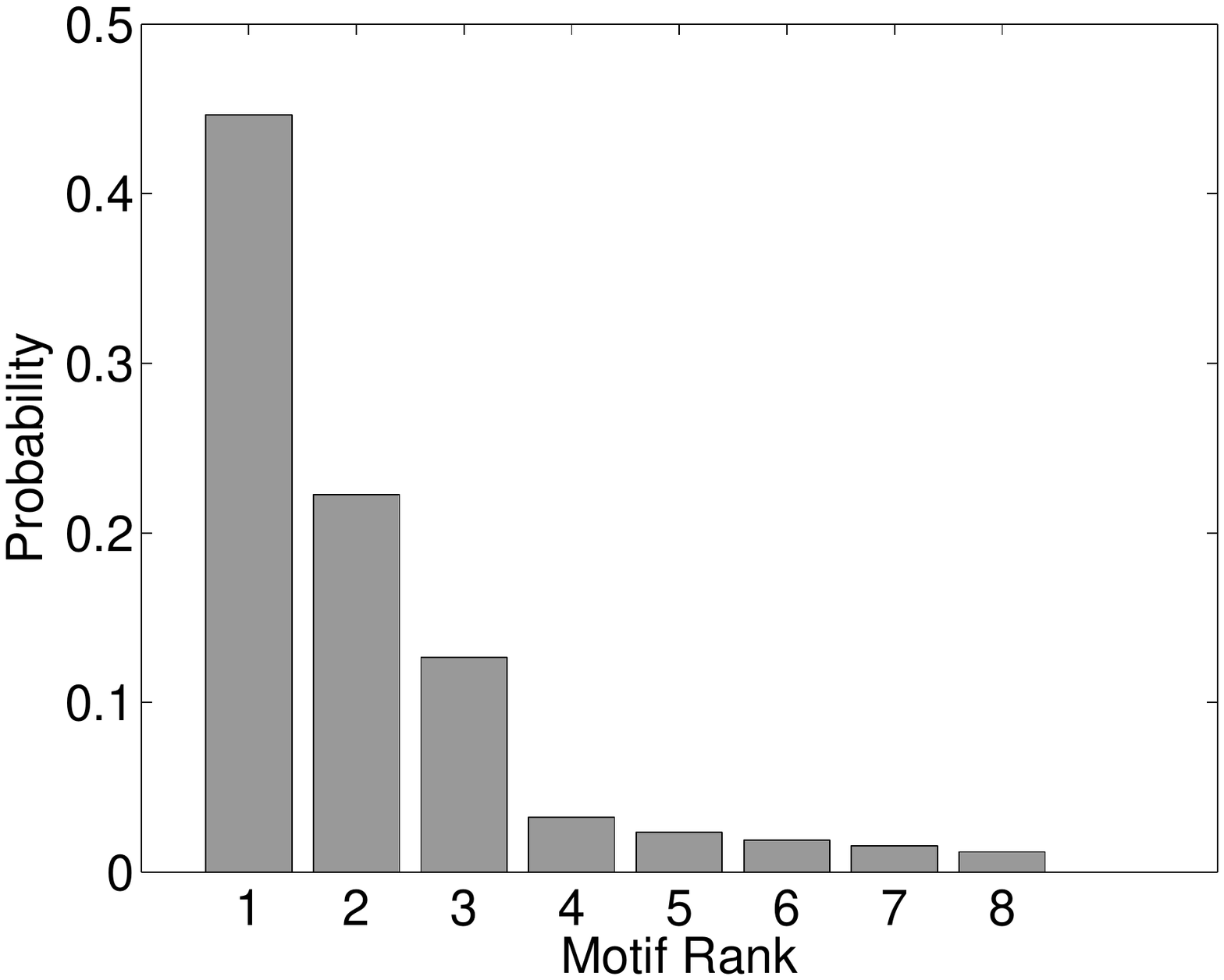}
				\put(17,15){
					\includegraphics[scale=.16,trim=0cm 0cm 0cm 0cm,clip]{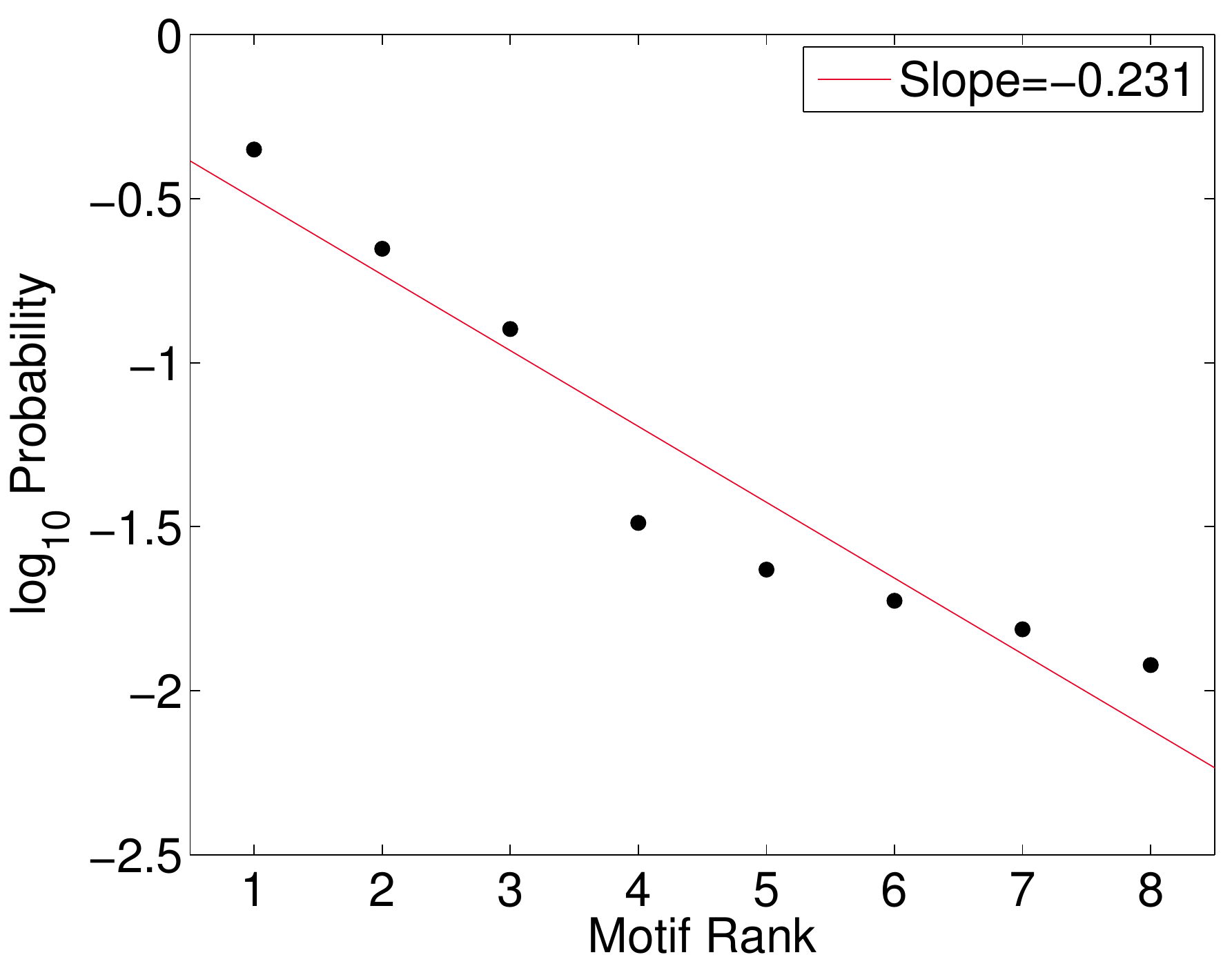}				
				}
			\end{overpic}\\ \hline
		\end{array}$
	\end{center}
	\caption{\bf We construct aggregated Twitter mobility motifs for each prolific user as directed networks, where tweet hubs for that user are represented by nodes, and a directed edge from node A to node B exists if a tweet was sent from tweet hub A and another tweet was sent from tweet hub B at most two hours apart. We separate each of these networks into their components and group these isomorphic subnetworks while counting how many of each transition between nodes occurs and the activity classification types of each of the nodes. We rank these subnetworks by occurrences among the prolific users and display the most common subnetworks in panels 1-8. The edges are labelled to convey the fraction of observed transitions represented by that edge. The nodes each have a subplot showing the probabilities that the tweet hub represented by that node has each of the activity classifications (H=Home, W=Work). The last panel shows the probabilities of these subnetworks in rank-order, along with a subplot containing the same distribution on a log-scale.}
	\label{motifs}
\end{figure*}

\indent We first investigate global statistics after identifying tweet hubs for each of the prolific Twitter users. Figure \ref{hubCount}A shows the probability an individual has $N$ tweet hubs. As was found in~\cite{motif}, we observe a trend that is approximately lognormal and is well modeled by 
\begin{equation}
	F(N;\mu,\sigma)=\frac{\exp\left(-\frac{(\ln N-\mu)^2}{2\sigma^2}\right)}{\sqrt{2\pi}N\sigma},
\end{equation}
where $\mu=1$, and $\sigma=1/2$; however, a slightly better fit in the least squares sense is obtained when $\mu=1.19$ and $\sigma=0.48$. This finding supports the ``universal law" proposed by Gonz\'alez et al. in~\cite{motif}. Figure \ref{hubCount}B demonstrates the relationship between ranking tweet hubs by tweet count and the probability that the individual sent a tweet from that hub. We provide a line with slope $-1$ as a guide for the eye. This linear trend may represent some underlying mechanism, but we do not endeavor to explore this further here.  \\
\indent We found that 68\% (65\% for English speakers) of tweet hubs in our dataset were classified as Home, while 32\% (35\% for English speakers) were classified as Work. Figure \ref{hubCount}C shows us how similar hourly PDFs for tweet hubs in our dataset were when compared to the activity class PDF (Fig.~\ref{BLScurves}) the tweet hub was classified as. We find that the root mean square error when comparing the hourly PDFs for tweet hubs and their activity classifications is almost always less than $0.4$, indicating that our classification method is performing reasonably well in terms of PDF similarity.\\
\indent We investigate the cumulative Twitter motifs in Fig.~\ref{motifs}, where we provide networks for each motif component that we encountered at least 1\% of the time. Each directed edge in these networks is labeled to show what fraction of transitions between tweet hubs in that component is explained by that edge. The nodes also have subplots displaying the probabilities for the tweet hub to have each of the activity classifications. The final plot shows the probabilities of observing each of the prominent motif components. \\
%
\indent We first notice that over 60\% of cumulative network components were found to be an isolated point or a 2-cycle (i.e. the Rank 1 or Rank 2 networks). Interestingly, one of the nodes in the 2-cycle is more likely to represent a Work tweet hub than the other and the edge originating from this node represents almost two-thirds of the transitions between these tweet hubs. This reflects the finding in~\cite{mystuff,hedono} that the volume of Twitter activity is increased in the evening when a working Twitter user would be commuting from work or school to home. \\
\indent We next note that the rank 4 through 8 motif components represent three hubs. The Rank 4, 5, 7, \& 8 networks suggest that nodes that are more likely to be Home than Work tend to have higher degrees, and the sum of weights from incoming edges tends to outweigh the sum of weights from outgoing edges. These observations suggest these nodes play a central role in the mobility patterns represented by these networks. \\
\begin{figure*}[!t]
	\centering
	\includegraphics[scale=.45,trim=0cm .5cm 1cm 0cm,clip]{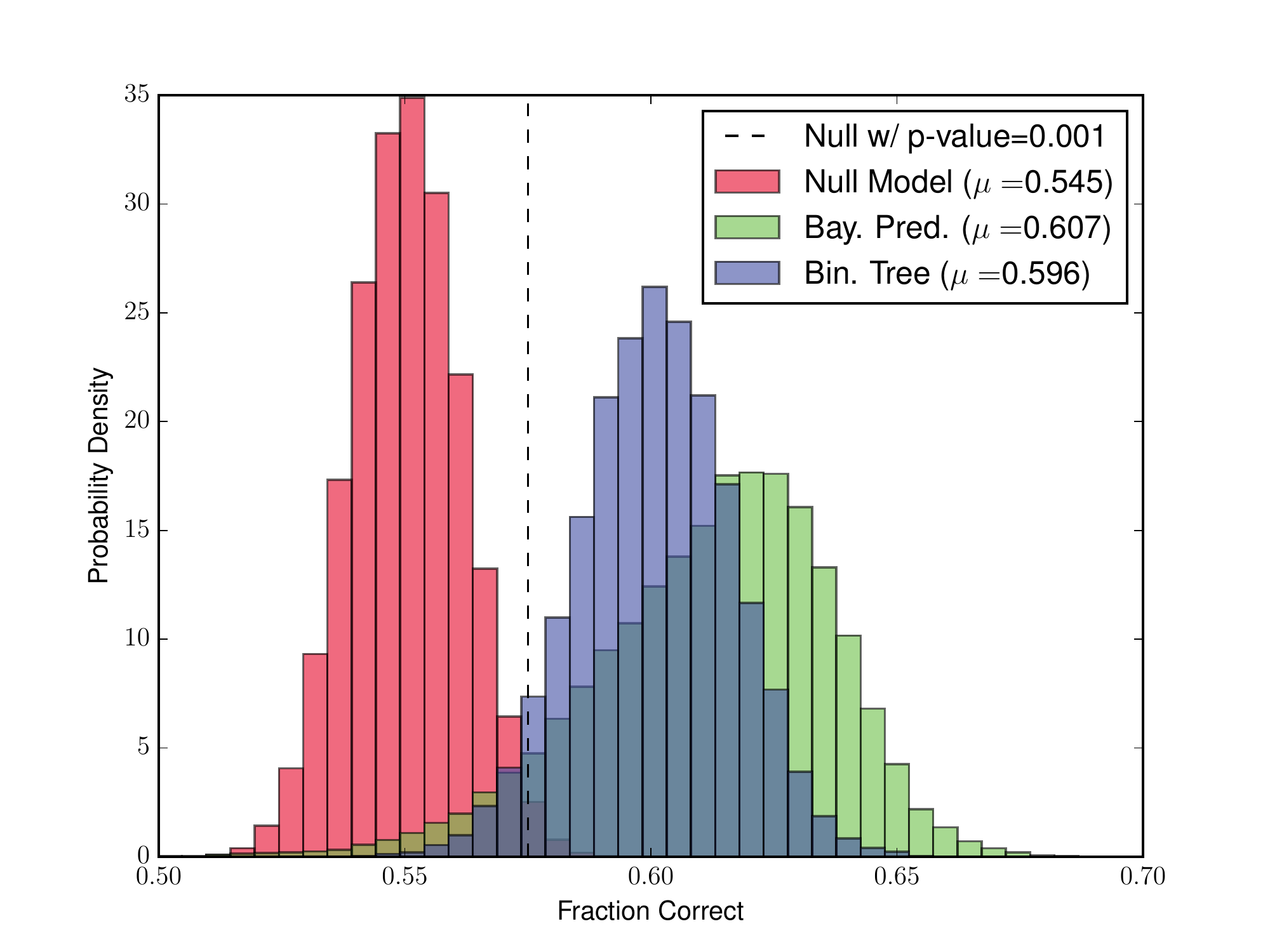}
	\caption{
		Distributions of the fraction of correct predictions made using three approaches: the null model (\textcolor{myRed}{red}); a pruned binary decision tree (\textcolor{myBlue}{blue}); and a Bayesian document classifier (\textcolor{myGreen}{green}). The latter two distributions are both estimated from 3,000 trials (see main text), and a ``fraction correct" 
calculated according to the number of correct predictions divided by the total number of tweet hubs in the validation dataset of that trial. The dashed line represents the value of the null model distribution above which has a p-value $<10^{-3}$.
	}
	\label{betterThanNull}
\end{figure*}
\indent We capitalize on the content of the Twitter messages by comparing binary decision trees and Bayesian document classifiers in predicting the activity classifications of tweet hubs based only on the vocabulary of the tweets contained in those hubs. In Fig.~\ref{betterThanNull}, we compare the fraction of correct classification predictions with the null model. Points to the right of the black vertical dashed line represent predictions that performed better than the null model distribution with a p-value$<10^{-3}$. For the binary decision tree and the Bayesian document classifier, we randomly select half of the prolific Twitter users in our dataset as training data and we record the percent of the correct activity classifications on the remaining half of the users as validation. Repeating this procedure for 3,000 trials, we obtain the blue and green distributions, representing the predictive power of the binary decision tree and the Bayesian document classifier, respectively. \\
\indent Both prediction methods yield a proportion of correct classification predictions that perform better than expected from the null model (i.e. p-value$<10^{-3}$). On average, the null model yielded correct predictions 54.5\% of the time, the binary decision tree yielded correct predictions 59.6\% of the time, and the Bayesian document classifier yielded correct predictions 60.7\% of the time. Throughout several trials of training and validating our predictors, we found that 15\% of Home tweet hubs and 20\% of Work tweet hubs were correctly classified at least 99\% of the time, indicating classification predictions were not random and not biased towards one class despite the different number of hubs of each class (see Fig.~\ref{hubCount}C). Furthermore, both of these predictive methods allow us to see which words contribute the most to the predictions. We provide the words that most strongly discriminate between Home (Fig.~\ref{revealWords}A) and Work (Fig.~\ref{revealWords}B). We see that pronouns (e.g. ``i", ``you", ``she"), slang (e.g. ``lol", ``lmfao", ``gonna"), and profanity (e.g. ``a**", ``b**ch") are more likely to occur from a Home tweet hub than from a Work tweet hub. On the other hand, references to time of day (e.g. ``lunch", ``afternoon", ``goodmorning"), references to weekdays, and references to work (e.g. ``business", ``working") are more likely to occur at Work tweet hubs. 
\section{Discussion}
\begin{figure*}[!Ht]
		\hspace{-.2cm}$\begin{array}{c}
			\begin{overpic}[width=\textwidth,trim=0cm 0cm 0cm 0cm,clip]{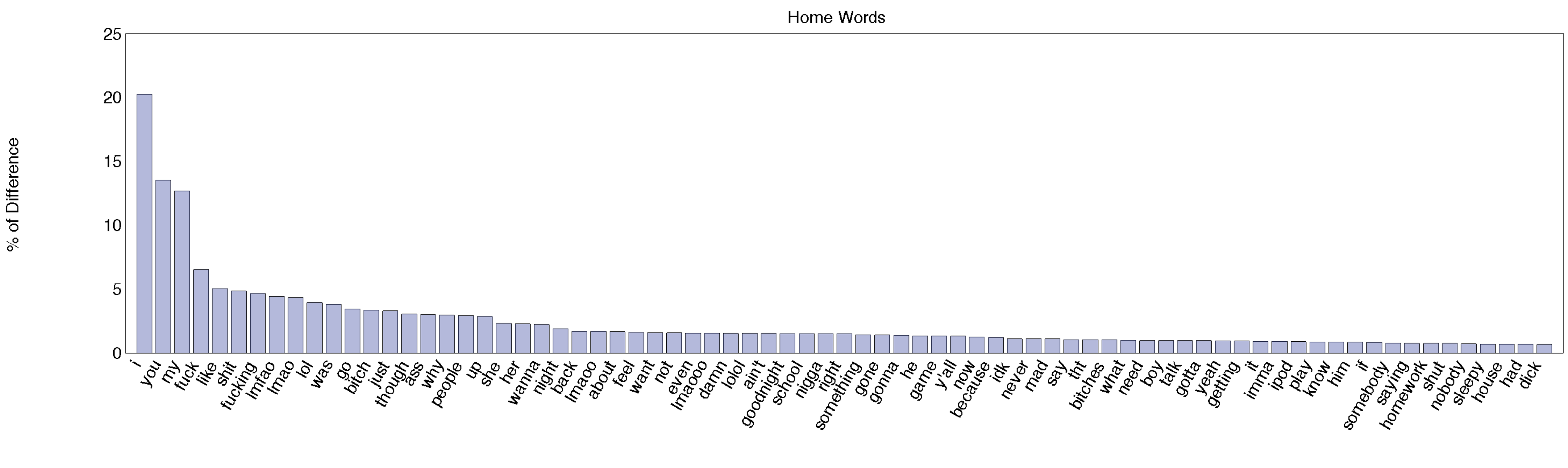}
				\put(20,40){\fbox{\text{A}}}
			\end{overpic}\\
			\begin{overpic}[width=\textwidth,trim=0cm 0cm 0cm 0cm,clip]{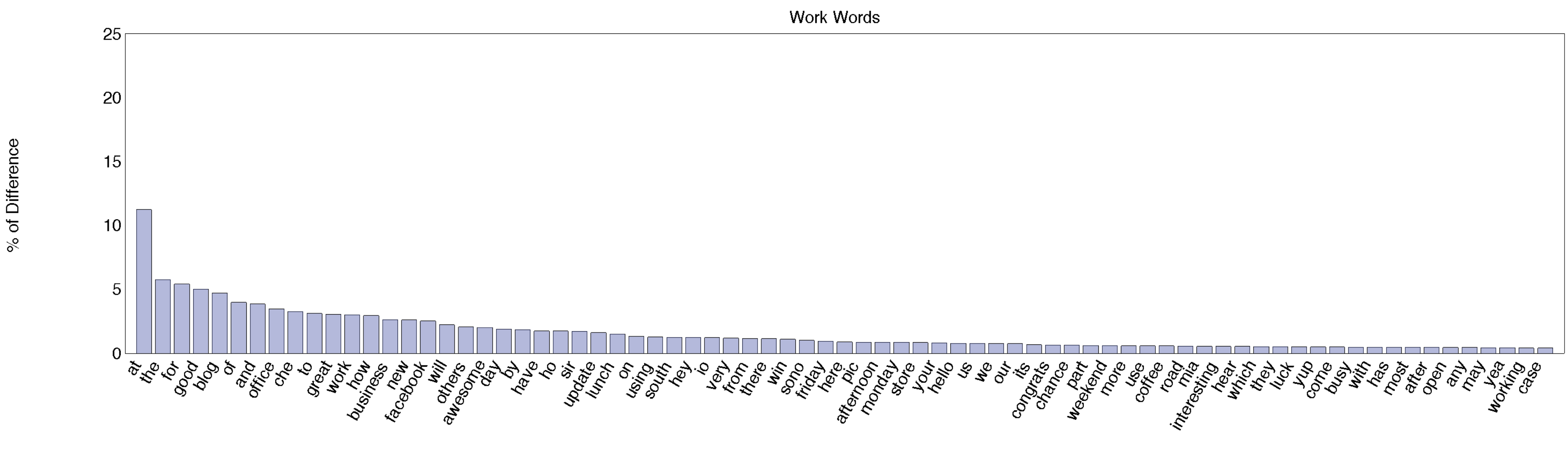}
				\put(20,40){\fbox{\text{B}}}
			\end{overpic}\\
	\end{array}$
	\caption{\bf Words ranked by their percent contributions to $P_{\mathrm{comp}}$ as terms in the sum from Eq.~(\ref{bayes3}) for tweet hubs that were consistently classified as Home (A), or Work (B). 
	}
	\label{revealWords}
\end{figure*}

\indent The purpose of this study is to use geolocated Twitter data to understand and characterize aspects of daily human life, such as identifying key spatial locations, how people travel between their locations, and why they visit those locations. Our work improves our understanding of sociotechnical systems in a way that may improve traffic models, models of disease spread, and current methods of urban planning by introducing methods for utilizing large-scale Twitter data that is abundant and publicly available. Using Twitter data for understanding sociotechnical systems grants us the interesting opportunity to use the content in the tweet messages to both asses our methods of characterizing human activity at key locations and infer the activities of individuals at locations with more incomplete data.\\
\indent We find evidence in our Twitter data to support observations on human mobility made through other data sources.
For example, Fig \ref{hubCount}A exhibits a lognormal distribution of tweet hub counts that agrees the finding of Gonz\'alez et al.~\cite{motif}, which employed both survey and cellphone data. Furthermore, we find agreement with previous studies finding that the mobility patterns of most people are explained by just three locations, and tweet hubs representing Home activity are more likely to play a centralized role in the mobility pattern (see Fig.~\ref{hubCount}A \& Fig.~\ref{motifs}). Finally, we find that we are more able to capture data from commutes from Work locations to Home locations, which may be related to previous findings regarding increased tweet volume in the latter portion of each day~\cite{mystuff,largeScale}. These conclusions may inform future studies by reducing the degrees of freedom required to model human mobility dynamics.\\
\indent We have introduced novel methods for observing characteristics of the daily lives of prolific Twitter users. Providing a way to identify key locations for individual users as tweet hubs allows us to analyze their mobility between those spatial locations, as well as attempt to understand why they visit those locations. We demonstrate our methods' ability to achieve the binary classification of tweet hubs between Home and Work by observing a notable subset of tweet hubs with activity classifications that are correctly identified almost all of the time, despite the difference in sample sizes among the Home and Work tweet hubs and regardless of what subset of our data we use to train our methods (see Fig.~\ref{betterThanNull}). Furthermore, an analysis of the distinguishing words for the two activity classes for the consistently correctly classified hubs supports our classifications by demonstrating the use of slang, profanity, and pronouns at Home, while tweets sent from Work hubs contained more formal grammar with references to work, time of the day, and day of the week (see Fig.~\ref{revealWords}).\\
\indent Finally, our use of the words contained in tweets from hubs introduces the exciting power of the content at a location to aid in identifying an individual's activity at that location. Fig.~\ref{revealWords} provides evidence that just the vocabulary of tweets sent from a location allows us to accurately predict the activities of the individual at that location. A more detailed understanding of language along with more sophisticated computational methods will allow researchers to identify the activities of an individual at a key location with greater sensitivity, beyond a binary classification.\\
%
\end{multicols}
\begin{multicols}{2}

\end{multicols}

\end{document}